\documentclass{emulateapj}

\usepackage{amsmath}
\usepackage[]{natbib}

\newcommand{\asec}{$^{\prime\prime}$}
\newcommand\hst{{\it HST}}

\slugcomment{Accepted for publication in the Astronomical Journal}

\shorttitle{The formation of massive galaxies: FW4871}
\shortauthors{Ferreras et al.}

\begin{document}

\title{The road to the red sequence: A detailed view of the formation of
a massive galaxy at z$\sim$2}


\author{Ignacio Ferreras\altaffilmark{1,2},Anna Pasquali\altaffilmark{3},
Sadegh Khochfar\altaffilmark{4}, Harald Kuntschner\altaffilmark{5},\\
Martin K\"ummel\altaffilmark{5}, Nor Pirzkal\altaffilmark{6},
Rogier Windhorst\altaffilmark{7}, Sangeeta Malhotra\altaffilmark{7}, 
James Rhoads\altaffilmark{7}, Robert W. O'Connell\altaffilmark{8},
Seth Cohen\altaffilmark{7} Nimish P. Hathi\altaffilmark{9},
Russell~E. Ryan, Jr.\altaffilmark{10}, \\
 Haojing Yan\altaffilmark{11}}
\altaffiltext{1}{ferreras@star.ucl.ac.uk}
\altaffiltext{2}{Mullard Space Science Laboratory, University College
  London, Holmbury St Mary, Dorking, Surrey RH5 6NT, UK.}
\altaffiltext{3}{Astronomisches Rechen Institut, Zentrum f\"ur
  Astronomie der Universit\"at Heidelberg, M\"onchhofstrasse 12--14,
  69120 Heidelberg, Germany}
\altaffiltext{4}{Theoretical Modelling of Cosmic Structures Group,
  Max-Planck-Institut f\"ur Extraterrestrische Physik,
  Giessenbachstr., D-85748 Garching, Germany}
\altaffiltext{5}{European Southern Observatory,
  Karl-Schwarzschild-Str. 2, D-85748, Garching, Germany}
\altaffiltext{6}{Space Telescope Science Institute, 3700 San Martin
  Drive, Baltimore, MD 21218}
\altaffiltext{7}{Department of Physics and Astronomy, Arizona State
  University, P.O. Box 871504, Tempe, AZ 85287-1504}
\altaffiltext{8}{Department of Astronomy, University of Virginia,
  Charlottesville, VA22904-4325}
\altaffiltext{9}{Observatories of the Carnegie Institute of
  Washington, Pasadena, CA91101}
\altaffiltext{10}{Physics Department, University of California, Davis,
  CA95616}
\altaffiltext{11}{Department of Physics and Astronomy, University of
  Missouri, Columbia, MO65211}

\begin{abstract}
Over half of the census of massive galaxies at z$\sim$2 are dominated
by quiescent stellar populations. The formation mechanism for these
galaxies is still under debate, with models relying either on massive
and early mergers, or cold accretion. It is therefore imperative to
understand in detail the properties of these galaxies. We present here
a detailed analysis of the star formation history (SFH) of FW4871, a
massive galaxy at z=$1.893\pm 0.002$.  We compare rest-frame optical
and NUV slitless grism spectra from the Hubble Space Telescope with a
large set of composite stellar populations to constrain the underlying
star formation history.  Even though the morphology features prominent
tidal tails, indicative of a recent merger, there is no sign of
on-going star formation within an aperture encircling one effective
radius, which corresponds to a physical extent of 2.6~kpc. 
A model assuming truncation of an otherwise constant SFH
gives a formation epoch z$_{\rm F}\sim$10 with a truncation after
2.7\,Gyr, giving a mass-weighted age of 1.5\,Gyr and a stellar mass of
0.8--3$\times 10^{11}$M$_\odot$ (the intervals representing the output
from different population synthesis models), implying star formation
rates of 30--110\,M$_\odot$/yr. A more complex model including a
recent burst of star formation places the age of the youngest
component at $145_{-70}^{+450}$\,Myr, with a mass contribution lower
than 20\%, and a maximum amount of dust reddening of
E(B--V)$<$0.4\,mag (95\% confidence levels). This low level of dust
reddening is consistent with the low emission observed at 24$\mu$m,
corresponding to rest-frame 8$\mu$m, where PAH emission should
contribute significantly if a strong formation episode were
present. The colour profile of FW4871 does not suggest a significant
radial trend in the properties of the stellar populations out to
3R$_{\rm e}$. We suggest that the recent merger that formed 
FW4871 is responsible for the quenching of its star formation.
\end{abstract}

\keywords{ Galaxies: elliptical and lenticular, cD -- Galaxies:
  individual: FW4871 -- Galaxies: stellar content}

\section{Introduction}

The formation of the most massive galaxies remains one of the main
challenges of galaxy formation models. Within the observed bimodality
of the galaxy population \citep[e.g.][]{kauf03,bal04}, massive
galaxies dominate the so-called red sequence, which feature mostly
old, passively evolving stellar populations, suggesting an early,
short-lived period of star formation.  The implied rapid and efficient
star formation history is corroborated by the presence of massive
galaxies at z$\sim$1-3 \citep{cim08,pg08} and by the mild evolution
with redshift of the corresponding comoving number density
\citep{fon06,con07,ig09,ban10}. On the other hand, the strong
evolution with redshift found on the mass-size plane
\citep[e.g.][]{dad05,truj06,lon07,vdk08} remains an open question,
with ongoing arguments debating whether the main size growth scenario
at z$\lesssim 2$ is due to either the emergence of early-type galaxies
with cosmic time, or to intrinsic mechanisms involving gas ejecta,
minor mergers, or major mergers
\citep[e.g.][]{ks06b,bour07,fan08,dam09,vdw09,naab09,shan10}, although
see \citet{rag11} and \citet{i3} for a criticism of the gas expulsion
scenario. The observed size-mass evolution of massive galaxies has
resulted in a major revision of our understanding of galaxy growth
\citep[see e.g.][]{na11}.

It is believed that two main channels of galaxy growth are important
under the hierarchical formation paradigm: one driven by shock heating
and subsequent cooling of the gas in halos \citep{wr78} and another
one driven by the accretion of cold gas along narrow filaments that
does not shock heat \citep{dekbir06,dek09,keres09}. At redshifts
z$\gtrsim$2, the presence of massive galaxies already in place suggests
that a very efficient mechanism must operate to fuel and transform
vast amounts of gas into stars within a dynamical time.

Wide, nearby surveys along with deep surveys complemented by high
resolution imaging are opening this important issue to scrutiny. In
the past, the properties of the stellar populations of z$\gtrsim$1
galaxies relied on photometric measurements or low SNR spectra, mainly
targeting the rest-frame NUV \citep[see e.g.][]{spin97}, with the
inherent degeneracies \citep{ig04}. The advent of the Wide Field
Camera 3 on board the Hubble Space Telescope has opened up an
important spectral window. With its IR slitless grisms (G102 and G141)
it is possible to target the Balmer break region of massive galaxies
over a redshift range (1.5$<$z$<$3) where our standard ideas of galaxy
growth can be put to the test.

We present here a detailed analysis of the star formation history of a
massive galaxy at z$\sim$2, discovered in the HST/WFC3 Early Release
Science by \citet{vdk}. The slitless grism data from WFC3-IR has
allowed us to probe in detail the star formation process of a massive
galaxy caught in its final merging stages towards the red
sequence. The age of the Universe at that redshift (around 3.4~Gyr)
enables us to constrain in more detail the age distribution. We
explore a large range of plausible star formation histories, with the
aim of estimating an upper bound to the recent and sustained star
formation rate for this galaxy, an issue that will shed light on the
main processes contributing to the growth of massive galaxies.  In
contrast to \citet{vdk}, our paper uses the latest flux calibration
available for the WFC3 grisms, a very important issue when extracting
star formation histories from spectral fitting. Furthermore, we
explore a huge volume of parameter space, instead of a few trial SFHs,
to determine not only the best fits, but robust uncertainties. In
addition, we compare two different sets of population synthesis models
to assess the systematics.

In this paper we assume a standard $\Lambda$CDM cosmology with 
H$_0$=70\,km/s/Mpc and $\Omega_m=0.3$. All magnitudes are given in the
AB photometric system.

\begin{figure}
  \includegraphics[width=9cm]{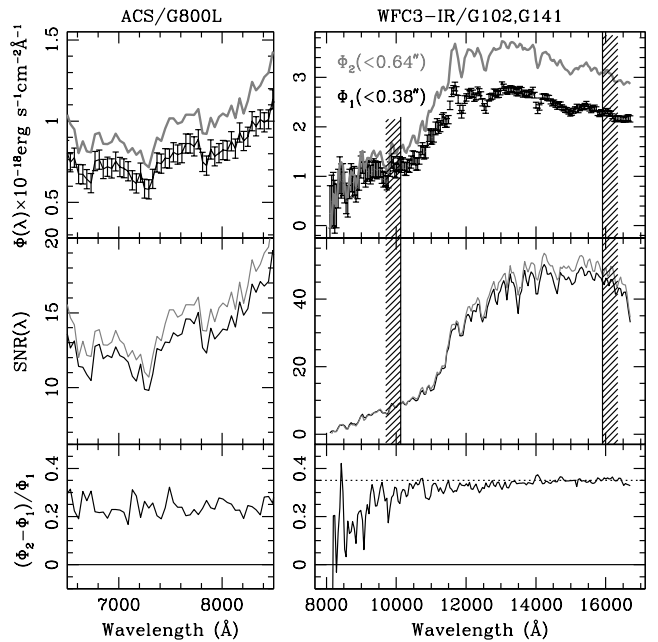}
 \caption{Slitless grism spectra of galaxy FW4871 from the Advanced
   Camera for Surveys ({\sl left}) and the Wide Field Camera 3-IR
   ({\sl right}). Two extractions are used in the analysis (wider
   extractions would include flux from a neighbouring object). The
   three panels, from top to bottom, show the flux, including error
   bars for the 0.38 arcsec extraction, SNR and flux ratio between the 
   extractions. The vertical shaded regions give the fitting interval
   for the analysis of the stellar populations.}
 \label{fig:seds}
\end{figure}

\section{Looking for passive galaxies with a slitless grism}

Following up our previous work on the spectral analysis of massive
galaxies at moderate-to-high redshift, we searched the WFC3/ERS
slitless grism dataset, looking for galaxies with a prominent Balmer
break. This technique has allowed us to constrain the star formation
history of early-type galaxies out to z$\sim$1.2 in the HUDF
\citep{grapes} and the wider HST/ACS coverage of the GOODS North and
South fields \citep{pears}. The IR coverage of WFC3 enables us to
probe the very sensitive region around the Balmer break at higher
redshifts.

We searched through the data available from the Early Release Science
programme (ERS) for Wide Field Camera 3 (WFC3), ID 11359 (PI
O'Connell). A field centred at RA = 03:32:17.6 and DEC = $-$27:42:32.4
(J2000) was imaged with an exposure time of 5017.6 s in each of the
WFC3/IR F098M, F125W and F160W filters. The data were reduced with the
standard WFC3 pipeline.  Slitless spectroscopy of the same field was
obtained through the WFC3/IR G102 (J band) and G141 (H band) grisms,
for a total exposure time of 4211.7 s per grism. The grism spectra
were reduced with the aXe 2.1 software\footnote{\tt
  http://axe.stsci.edu/axe/index.html} \citep{mk11}.  The data cover
a field of view of 4.65~arcmin$^2$ \citep[see][for details]{stra11}
with a spectral coverage of 0.8--1.6$\mu$m and resolution between
R=210 at $\lambda=1\mu$m and R=130 at 1.4$\mu$m. The spectral
coverage allows us to probe the Balmer break in the redshift window
1$<$z$<$3, although the shallowness of the ERS data restricts this
range to $z\lesssim 2$ for the case of massive (M$_s\gtrsim
10^{11}$M$_\odot$) galaxies, where M$_s$ is the stellar mass.

From this dataset, we found only one galaxy with a high enough S/N for
the application of our methodology to extract star formation
histories. This galaxy has been already presented in \citet{vdk}, and
identified as FW4871 from the FIREWORKS catalogue \citep{fw}. In this
paper we look in detail at the constraints one can impose on its
stellar populations, employing a vast range of star formation
histories.

\begin{figure*}
\begin{center}
\includegraphics[width=14cm]{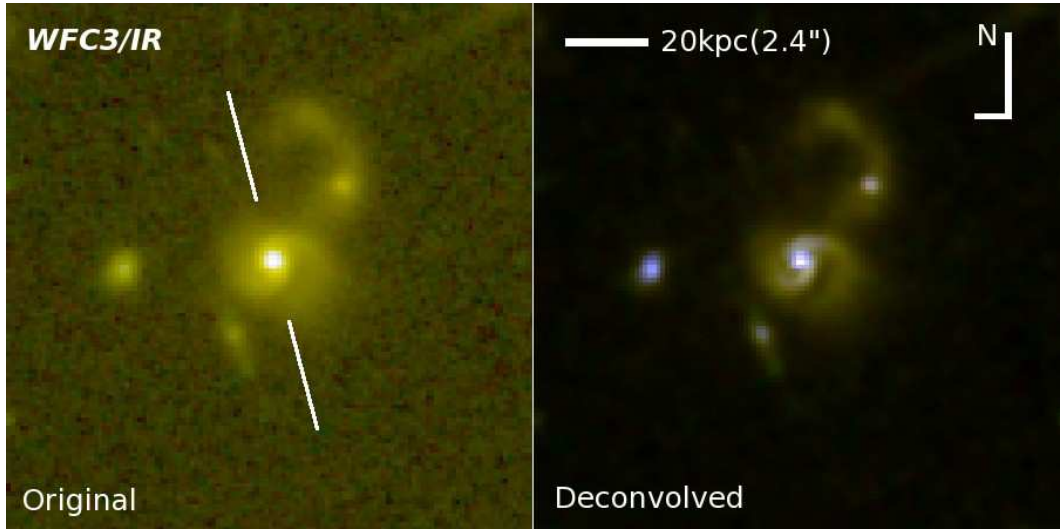}
  \caption{Colour montage of FW4871 using WFC3/IR images in the F098M
    (blue), F125W (green) and F160W (red) passbands. The panels
    illustrate the effect of the deconvolution. Notice the nearby
    distorted galaxy located NW of FW4871 is a companion
    \citep[identified as FW4887, ][]{vdk}. The orientation of the
    extraction of the spectrum is shown as two white segments in the
    left-hand panel.}
 \label{fig:clr}
\end{center}
\end{figure*}

\begin{deluxetable*}{ccccccccc}
\tabletypesize{\scriptsize}
\tablecaption{Observations (AB magnitudes, errors given at the $1\sigma$ level)}
\tablewidth{0pt}
\tablehead{\colhead{Redshift} & \colhead{F160W$^1$} & \colhead{R$_e$($^{\prime\prime}$)} & 
\colhead{n$_S$} & \colhead{Aperture} & \colhead{F098M--F160W$^2$} & \colhead{F125W--F160W$^2$} & 
\colhead{MgUV$^{2,3}$} & \colhead{D4000$^4$}}
\startdata
1.893$\pm$0.002 & 19.81$\pm$ 0.07 & 0.31$\pm$ 0.04 & 5.4$\pm$ 1.1 & 0.38$^{\prime\prime}$ & 1.89$\pm$0.34 & 0.35$\pm$0.22  & 1.16$\pm$0.04 & 1.314$\pm$0.013\\
  &   &   &   & 0.64$^{\prime\prime}$ & 1.88$\pm$0.26 & 0.40$\pm$0.16 & 1.15$\pm$0.05 & 1.297$\pm$0.012\\
\enddata
\tablenotetext{1}{Total apparent magnitude corresponding to the S\'ersic fit.}
\tablenotetext{2}{Measured within the apertures (given as diameters).}
\tablenotetext{3}{As defined in \citet{dad05}.}
\tablenotetext{4}{As defined in \citet{bal99}.}
\label{tab:Obs}
\end{deluxetable*}

To further constrain the models, we also extracted the observed-frame
optical spectrum of FW4871 from the 2005 observations with the
Advanced Camera for Surveys (ACS) through its WFC/G800L grism, as part
of the program Probing Evolution And Reionization Spectroscopically
(PEARS, ID 10530, PI Malhotra). The total exposure time was 15140s,
and the spectrum was extracted and calibrated with the aXe software
\citep{mk09}.

In order to avoid contamination from a nearby source in the grism
images, we have extracted two optical$+$IR spectra for FW4871, over
apertures of three and five pixels on the WFC3-IR grism data,
corresponding to diameters of 0.38 and 0.64~arcsec, respectively. The
IR G102 spectra have been degraded to the same spectral resolution of
those acquired with G141 (R = 130 at 1.4 $\mu$m). The original
spectral resolution of the ACS spectra is preserved, since these data
will be fitted independently of the IR data. Fig.~\ref{fig:seds} shows
both ACS and WFC3 spectra within these two apertures. The S/N (mid
panels) is high enough for a detailed analysis of the stellar
populations (\S4).  Notice the lack of strong emission lines ([\ion{O}{2}],
H$\beta$, and [\ion{O}{3}] fall within the wavelength coverage of the WFC3
data) and the presence of prominent Balmer lines, typical of stellar
populations that underwent a recent period of star formation, i.e. a
k+a galaxy \citep{dg92}.

The spectral fitting method requires knowledge of the effective
spectral resolution. We matched a set of simple stellar populations
with similar absorption lines (around an age of 1\,Gyr) against the
data, and smoothed the spectra using a Gaussian kernel with variable
width. We also explored a range of redshifts for an accurate estimate.
We refined the redshift estimate from z=1.902 in \citet{vdk} to
z=1.893, and obtained an effective resolution FWHM of 120\AA, in
agreement with the expected resolution of G141 (R=130 at 1.4$\mu$m).
We note that the spectral resolution of slitless grism spectroscopy is
strongly affected by object size. The galaxy presented in this paper
is, of course, resolved, which implies a lower spectral resolution
than the camera specification values, given for unresolved sources.

\begin{figure}
  \includegraphics[width=9cm]{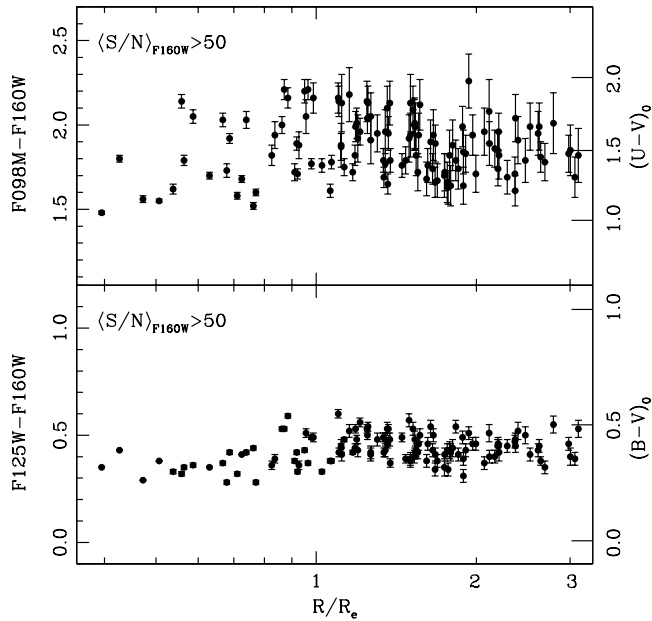}
 \caption{Colour profile after a Voronoi tessellation of the WFC3/IR
   images. The tessellation is performed with a target S/N per bin
   in the F160W passband above 50. The right axes give the
   K-corrected, rest-frame U--V (top) and B--V colours.}
  \label{fig:cgrad}
\end{figure}

\section{Intrinsic colour profile}

In addition to the spectroscopic data, we make use of the NIR images
from WFC3/ERS \citep{ers}. Imaging through three passbands are
available: F098M, F125W and F160W, reaching a 50\% point-source
completeness limits of AB(F098M)=27.2 mag, AB(F125W)=27.5 mag, and
AB(F160W)=27.2 mag, at the 5$\sigma$ level \citep[][]{ers}. At the
redshift of our target, these three passbands closely map rest-frame
$U$, $B$ and $V$ standard filters. We ran a set of simple stellar
populations from the latest versions of the \citet{bc03} models to
determine the K-corrections relating these observed and rest-frame
filters. As expected, the K-corrections are rather low, always below
0.05~mag even if the metallicity is unknown over one decade.

\begin{figure*}
\begin{center}
  \includegraphics[width=8cm]{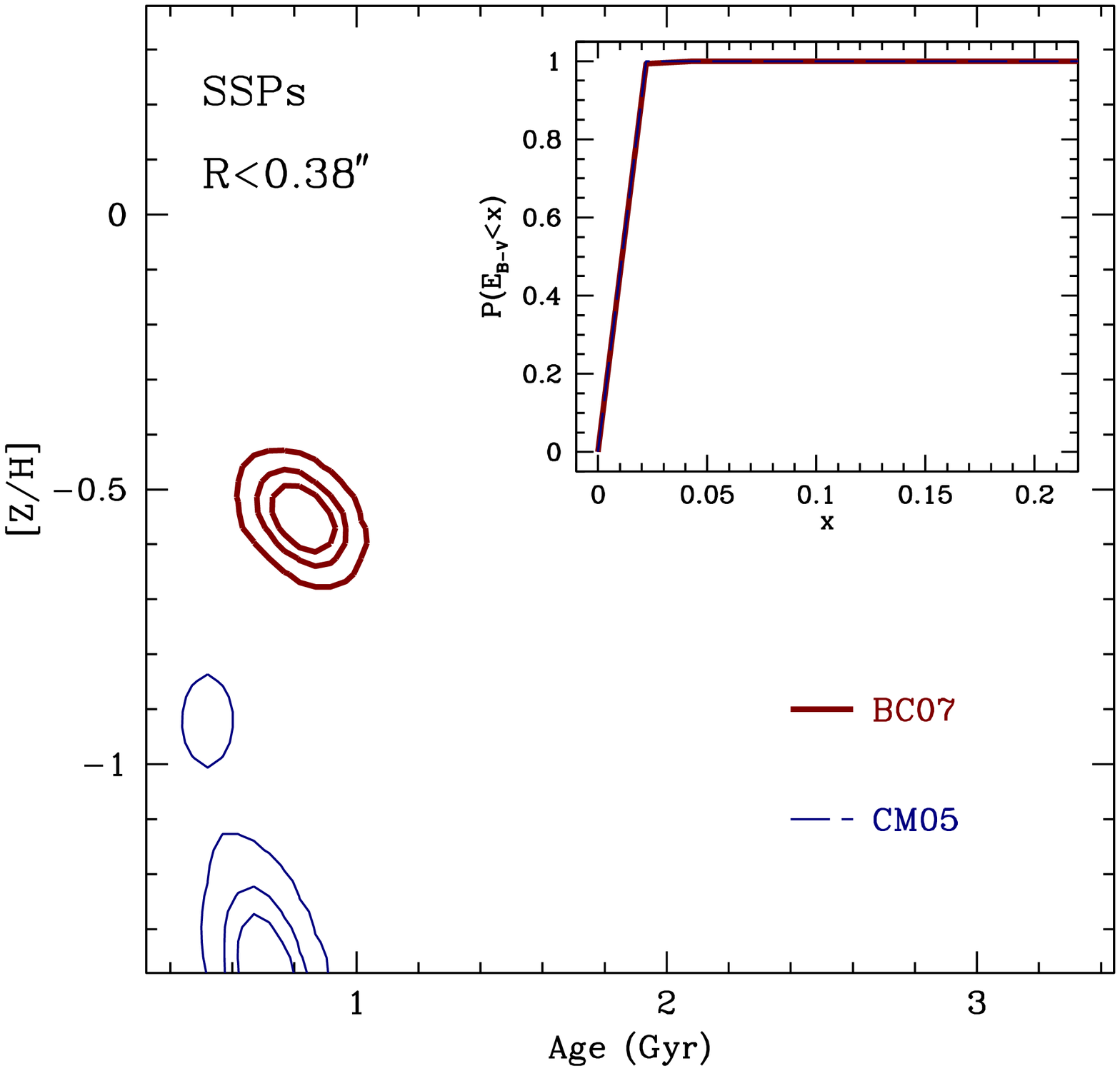}
  \includegraphics[width=8cm]{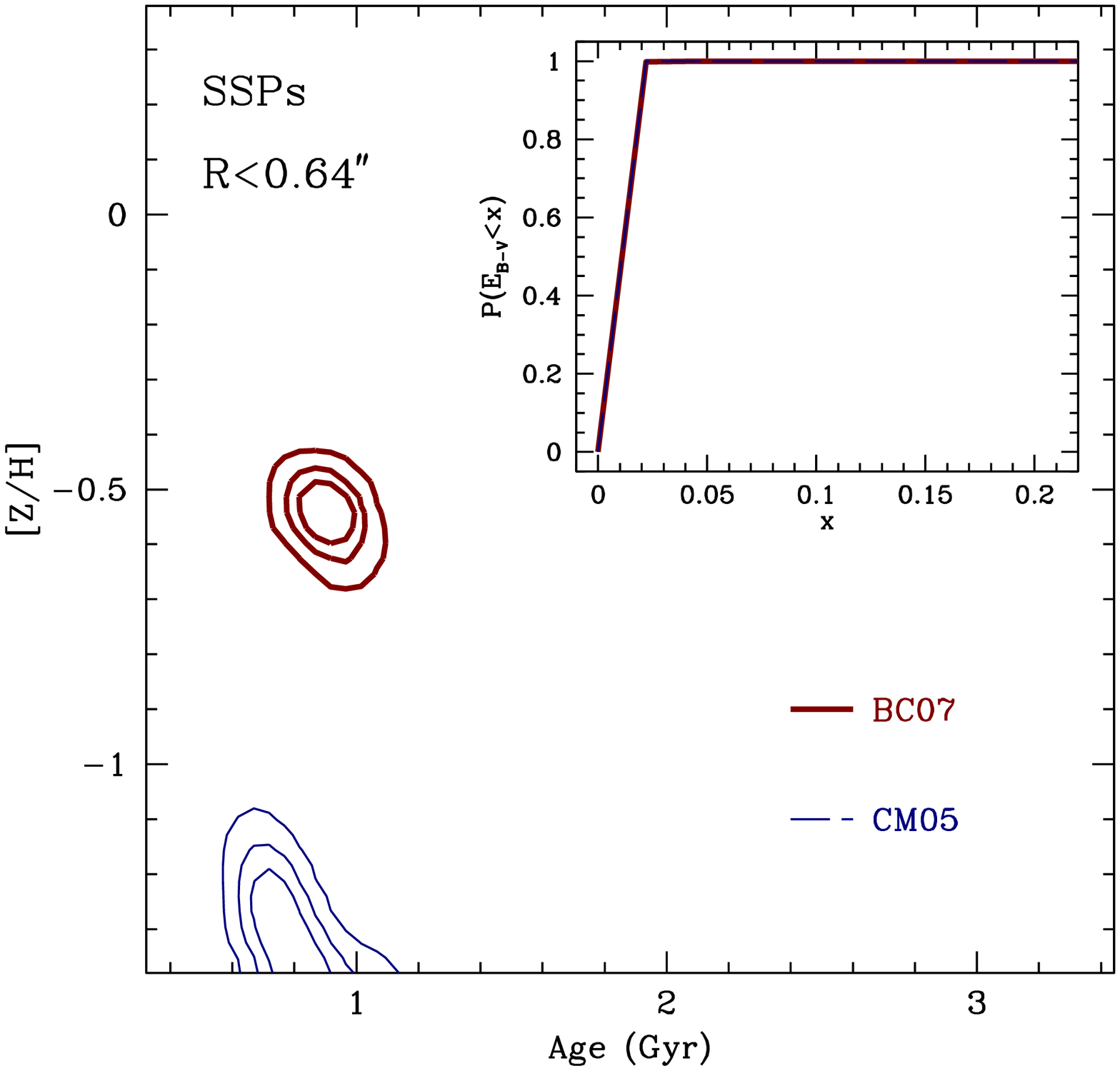}
  \caption{The contours show the 69, 90 and 99\% confidence levels for
    the spectral fitting of the WFC3 (rest-frame optical) and the ACS
    data (rest-frame NUV). In this case the models used are Simple
    Stellar Populations (i.e. a single age and metallicity), including
    dust as a free parameter. The inset shows the cumulative
    distribution of dust. We compare the 2007 version of the
    \citet{bc03} models with the models of \citet{cm05}, as
    labelled. Two SED extractions of the WFC3 data are considered: a 3
    pixel one ({\sl Left}, labelled R$<$0.38\asec, corresponding to a
    physical extent of 3.2~kpc) and a 5 pixel extraction ({\sl Right},
    labelled R$<$0.64\asec, corresponding to 5.4~kpc).}
  \label{fig:ssp}
\end{center}
\end{figure*}

In order to constrain the surface brightness profile of the galaxy, we
ran GALFIT \citep{galfit} on the three WFC3/IR images, for a Sersic
profile plus a constant sky background.  The PSF is built from a set
of 11 stars extracted from the same images. The results are given in
Tab.~\ref{tab:Obs}, where the uncertainties are quoted at the
$1\sigma$ level and are computed from a set of GALFIT runs on the same
images, replacing the combined image of the PSF by those from the
individual stars. The half-light radius (0.31$\pm$0.04 arcsec) maps
into a projected physical distance of 2.64$\pm$0.35~kpc at the
redshift of FW4871, which makes this object a typical example of the
compact massive galaxies found at similar redshift
\citep[e.g.][]{dad05,truj06,vdk08}, although we note that the CM05
models would locate this galaxy closer to the local sample on the
mass-size plane (see below). The two extractions of the spectra we
will use for the analysis roughly correspond to apertures with
diameter 0.38\asec, and 0.64\asec, hence enclosing R$_e$/2 and R$_e$,
respectively. The high value of the S\'ersic index makes the surface
brightness profile compatible with an early-type morphology.

In order to determine the intrinsic colour gradient of FW4871, we
follow the approach described in \citet{ig05} involving a
Voronoi tessellation to increase the S/N of the image, preserving as
much as possible the spatial information.  
Fig.~\ref{fig:clr} shows the original frames ({\sl left}), and a
deconvolved reconstruction ({\sl right}), using the {\sl lucy} task in
IRAF \citep{lucy74}. The figure uses all three NIR passbands to
generate an RGB colour image. This galaxy reveals itself as a major
merging event, with nearby objects displaying tidal features as well,
most notably the galaxy to the NW, at the same redshift, identified as
FW4887, a galaxy with a mass $\sim$1/10 of FW4871, which will likely
contribute to a later minor merger with FW4871 \citep{vdk}. The
spiral-like shape of the tails might alternatively suggest a late-type
galaxy morphology. However, as we will show below, the lack of dust
and ongoing star formation from the analysis of the spectra makes this
option unlikely.

The Voronoi tessellation targeted a S/N of 50 in F160W, including all
pixels within 3R$_{\rm e}$. For instance, in the F098M--F160W colour
profile, out of 1093 pixels we get 128 binned tiles.
Fig.~\ref{fig:cgrad} shows both colour profiles from the WFC3/IR data,
with the corresponding rest-frame U--V and B--V colours on the right
axes (the K-corrections were computed for a range of simple stellar
populations at solar metallicity).
We note that an analysis of the deconvolved images give no significant
difference, confirming that the change with respect to the PSF for the
three WFC3/IR passbands used, introduces no significant biases in the
colors. Within the core (R$\lesssim$R$_e$) the colours are slightly
bluer than in the outer regions, but still around (U--V)$_0\sim$1,
reflecting either small levels of ongoing star formation or dust
reddening. The following section presents the spectroscopic analysis,
that provides more stringent constraints on the underlying stellar
populations.

\section{Modelling the star formation history}

The high S/N of the WFC3 IR slitless grism data allows us to perform
the fitting of the spectral energy distribution corresponding to the
rest-frame 3500--5500\AA, i.e. a very sensitive region to the age
and metallicity distribution. We note that a contaminant spike is present
in the original WFC3 SED at an observed $\lambda$=15100\AA. At the
redshift of the galaxy, this contamination appears very
close to the metallicity-sensitive MgI region around 5170\AA\
\citep[see fig.~3 of ][]{vdk}. In order to be able to use this
region, we corrected it by running the CST models (described below),
fitting the spectrum over the wavelength range 3500--5100\AA. The
best fit (extended to 5500\AA ) is then used to perform a correction
in the region of the spike -- which extends over $\sim 200$\AA,
redward of the MgI feature. We tested our results in the two cases,
i.e. without any correction, fitting only blueward of 5100\AA, and
with the correction, fitting out to 5500\AA, to find no significant
difference in the age distributions, but an important suppression of
the probability distribution towards very low metallicities.

The spectrum (shown in Fig.~\ref{fig:seds}) shows prominent Balmer
lines, characteristic of stellar populations of ages around 1~Gyr. We
used the ACS data to measure the MgUV feature (a local peak in the
heavily-blanketed NUV continuum at rest-frame 2625-2725\AA) and
obtained a value of MgUV=$1.16\pm 0.04$, which is also indicative of a
similar age \citep[see e.g. fig.~2 in][]{dad05}. Our aim in this
section is to characterize the past star formation history of
FW4871. The discovery paper of this system \citep{vdk} already
presented a first approach to this problem, using the method outlined
in \citet{kriek09}. In this paper we explore in more detail this
issue, including additional information from the MgUV feature and
determine robust uncertainties on the derived formation history by the
use of a large number of models. We ran four separate grids of
models, as outlined below. For each choice of the star formation
history, we define a likelihood combining -- as independent
observables -- the WFC3 spectral data, the measured colours
(F098M--F1609W and F125W--F160W), and the MgUV index obtained from the
ACS spectrum. Notice we have two separate runs for a WFC3 extraction
of 3 and 5 pixels, corresponding to diameters of 0.38 and 0.64 arcsec,
respectively. The ACS and photometric data are extracted to match the
same apertures (see Tab.~\ref{tab:Obs} for the values used).  The
best fits obtained for these models are shown in
Tab.~\ref{tab:ModelPreds}.

\begin{figure*}
\begin{center}
  \includegraphics[width=8cm]{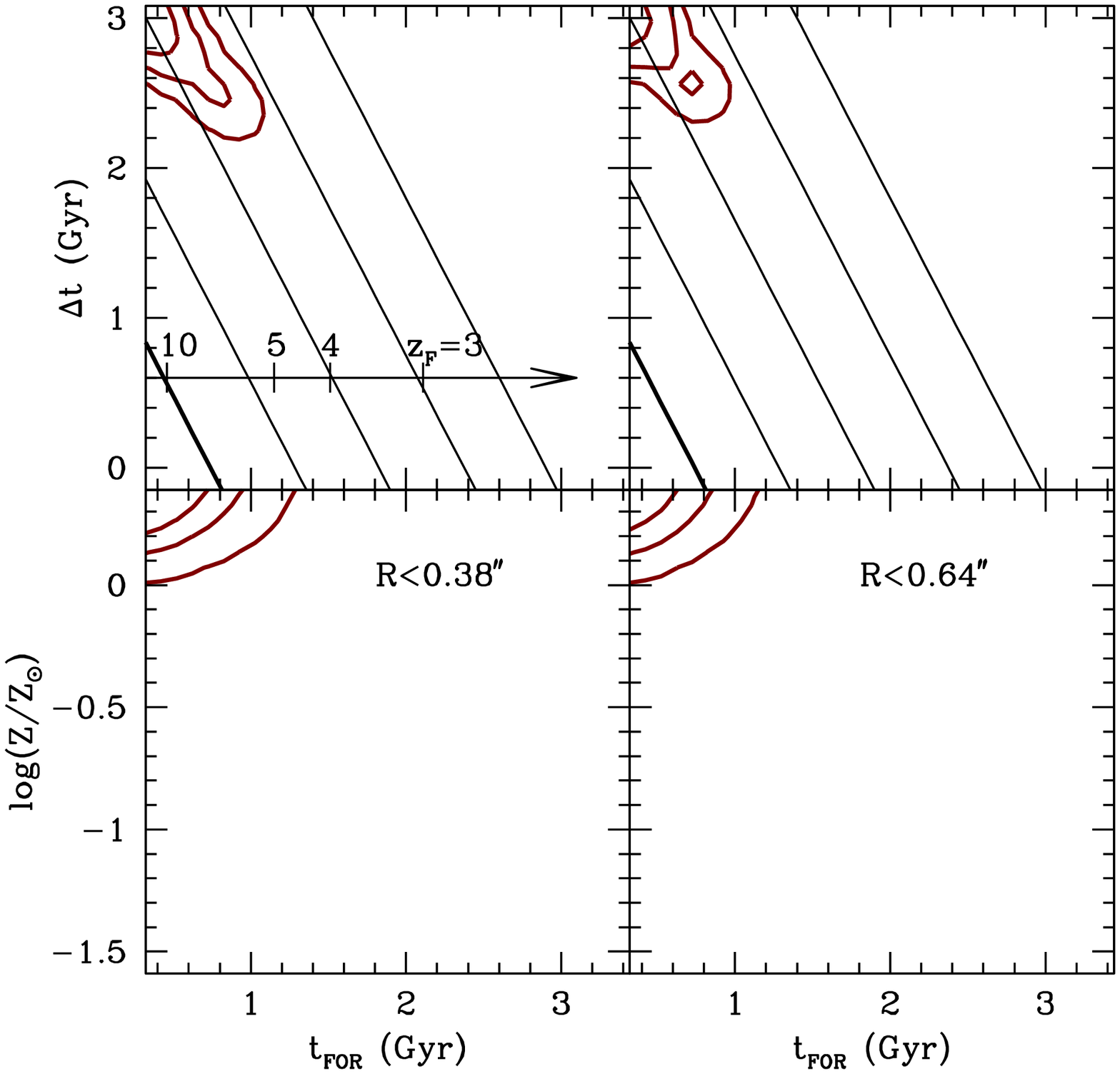}
  \includegraphics[width=8cm]{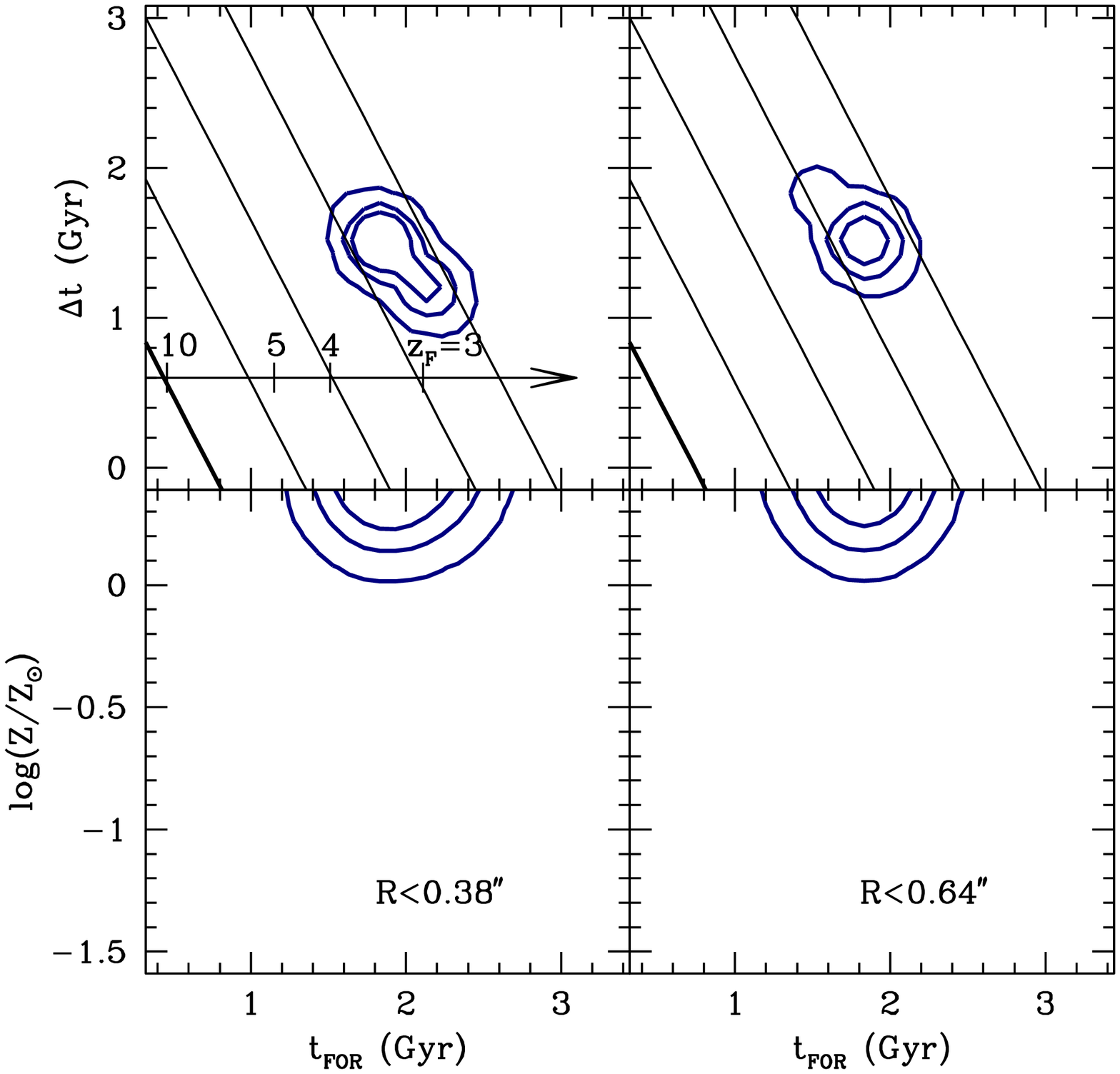}
  \caption{As in Fig.~\ref{fig:ssp} for the fit from the truncated
    models (CST, see text for details). The panels on the left (right)
    correspond to the BC07 (CM05) models. The top panels also show the
    contours for an mass-weighted average age from 0.5~Gyr to 2.5~Gyr
    (thick line) in steps of 0.5~Gyr. The composite models are based
    on the population synthesis models of BC07 ({\sl left}) and CM05
    ({\sl right}).}
  \label{fig:cst}
\end{center}
\end{figure*}

\begin{figure*}
\begin{center}
  \includegraphics[width=8cm]{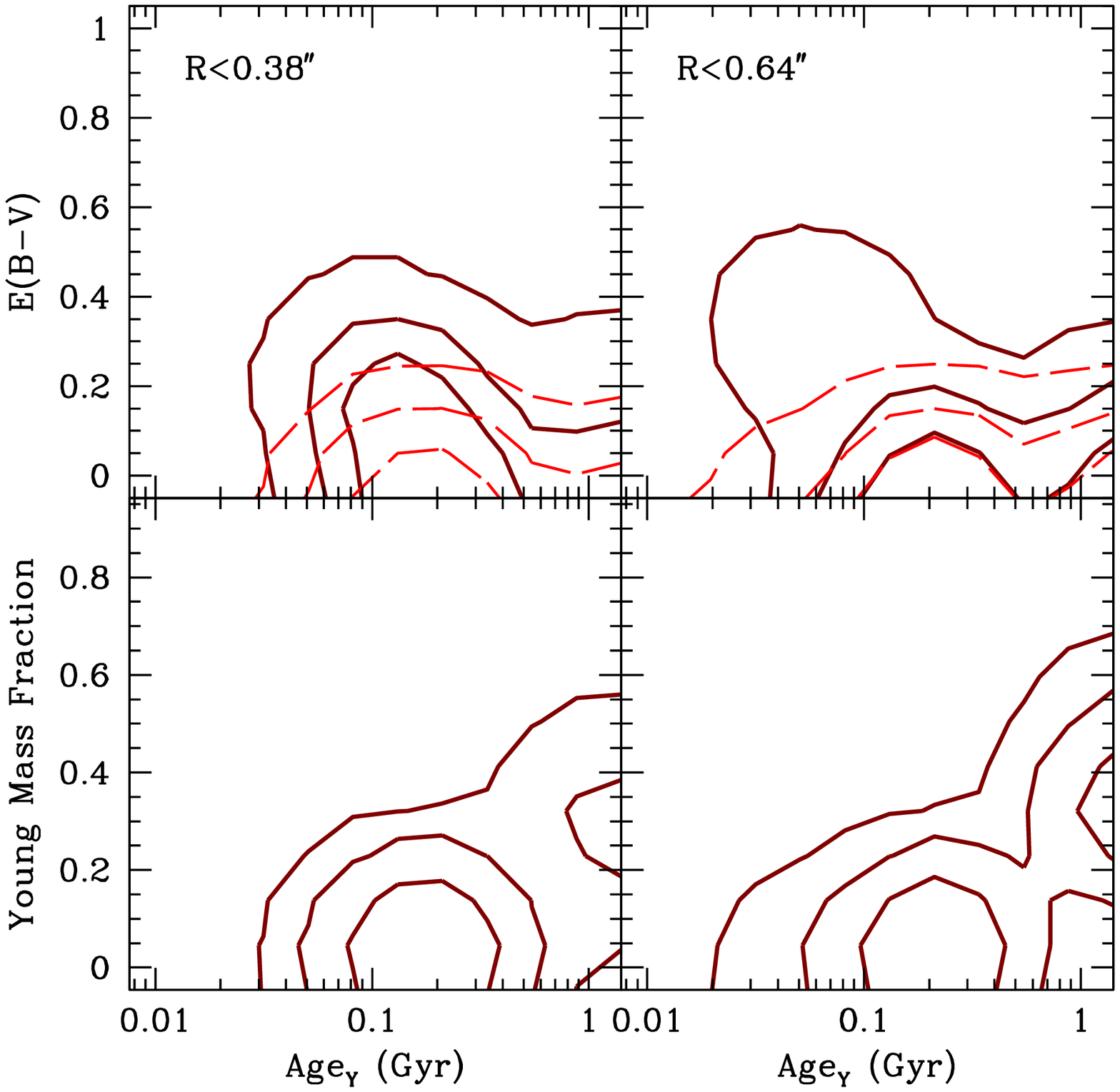}
  \includegraphics[width=8cm]{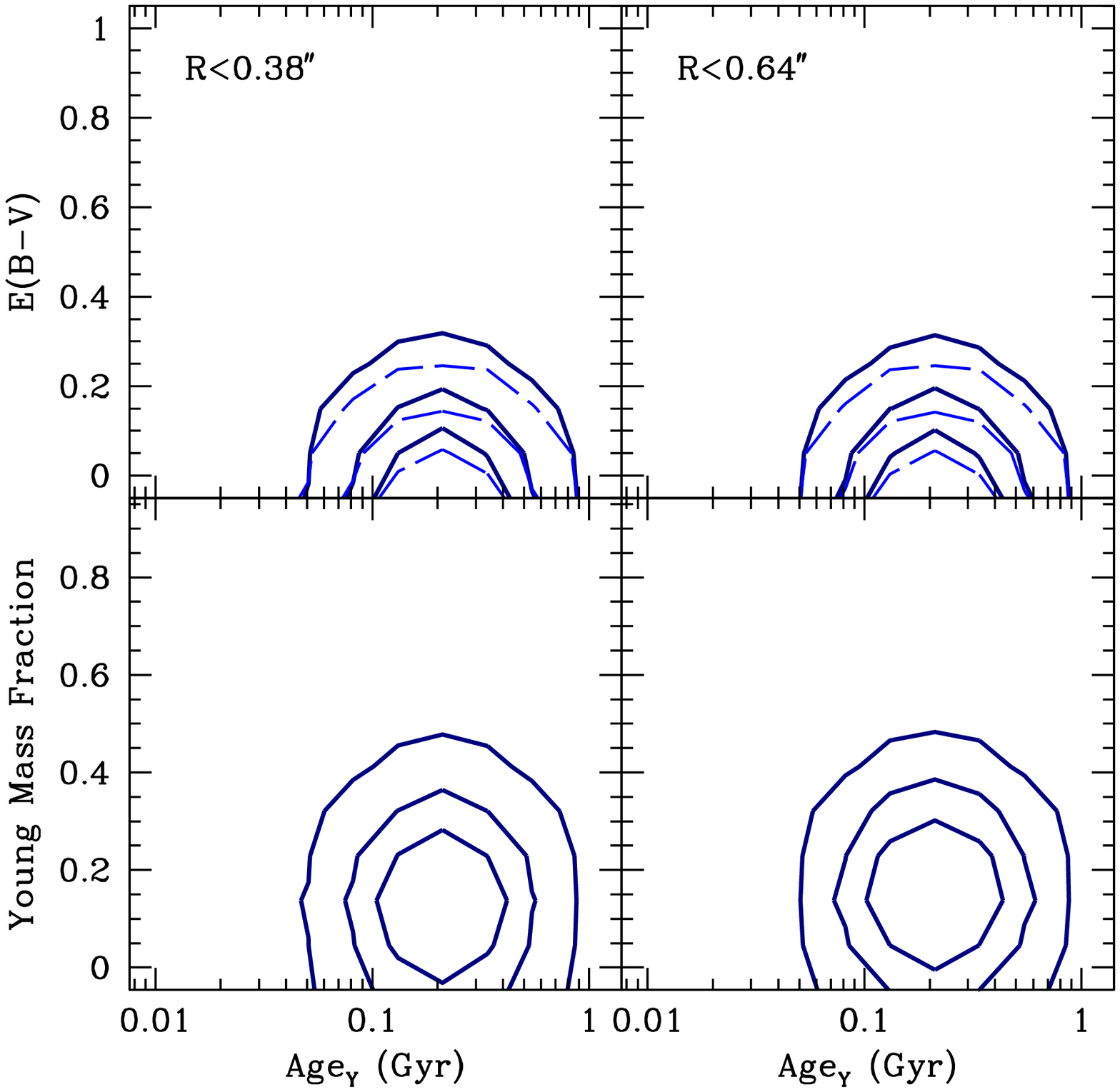}
  \caption{Contours are shown at the 69, 90 and 99\% confidence levels
    for the fit corresponding to the truncated models plus a later
    burst in the form of a simple stellar population (CST2, see text
    for details).  The figure shows the properties of the secondary
    (younger) component, as labelled. The older component features
    similar constraints to the CST models shown above
    (i.e. Fig.~\ref{fig:cst}). In the top panels we show the
    constraints for the independent dust reddening parameters of the
    old (dashed lines) and the young (solid lines) components.  The
    composite models are based on the population synthesis models of
    BC07 ({\sl left}) and CM05 ({\sl right}).}
  \label{fig:cst2}
\end{center}
\end{figure*}

\subsection{Simple Stellar Populations}

A first approach would involve simple stellar populations (SSP),
defined by a single age and metallicity. We also include dust as a
free parameter described by E(B--V), modifying the spectrum with two
reddening laws: the \citet{calz01} law -- appropriate for starbursting
systems -- and a Milky Way reddening law \citep{fitz99}.  The grid of
model parameters is defined in
Tab.~\ref{tab:Params}. Fig.~\ref{fig:ssp} shows the confidence
levels on age and metallicity, marginalized over dust, for two
population synthesis models: the 2007 version of the \citet{bc03}
models \citep[BC07; ][]{bruz07}, and those of \citet[][hereafter
  CM05]{cm05}. Each panel in Fig.~\ref{fig:ssp} corresponds to one
of the two extractions, as defined above (other results are
shown in Tab.~\ref{tab:ModelPreds}, with the uncertainties given at
the 5 and 95 percentiles).  Note the significant difference with
respect to stellar mass between BC07 and CM05 models, a well-known
feature caused by the different prescriptions included in these
models, most notably the modelling of the TP-AGB phase\citep{cm06}.
The insets show the cumulative probability distribution
of dust, which shows good agreement between both models, with values
below E(B--V)=0.02 at the 95\% confidence level. Given the merger-like
morphology of the object (Fig.~\ref{fig:clr}) it is quite remarkable
to obtain such a low level of dust and a relatively ``old'' SSP-age
(the presence of ongoing star formation would shift the derived age to
lower values). However, it is a well-known fact that SSP-based fits
are not robust indicators of age unless we are dealing with a system
formed over a short period of time compared to stellar evolution
timescales (e.g. a globular cluster), or for systems with old and
early-truncated star formation histories \citep[e.g. a massive
  early-type galaxy, see][]{bmc}. Notice that the best fit SSP gives a
rather low metallicity, especially for the CM05 models. The constraint
on a single age to explain a composite population can result in such
values on the metallicity. To overcome this, we consider more
realistic composite models where the star formation history is
parameterised by a small number of variables. A different reddening
law does not change the outcome, a result that will be consistent for
the composite models. The reason is that the reddening law used in
this analysis pertains mainly to the optical region, where most
reddening laws do not differ much. In the NUV, the only information
used -- the MgUV index -- is rather immune to a change in reddening
law given that it is defined over a small spectral window, and it is
located away from the ``2175\AA\ bump'' \citep[see e.g.][]{bump}.

\begin{deluxetable*}{llcrrrrrr}
\tabletypesize{\scriptsize}
\tablecaption{Parameter fits, combining the WFC3/IR SED, the MgUV
    index (measured from the ACS SED) and F098M, F125W and F160W
    photometry. Ages are defined as mass-weighted values. For the EXP
    models, the age dispersion $\Delta t$ is defined as the RMS of the
    distribution, for CST models it represents the duration of the
    burst. Error bars are quoted at the 5--95 percentiles.}
\tablewidth{0pt}
\tablehead{\colhead{Model} & \colhead{PopSyn} & \colhead{Aperture} & \colhead{$\langle$Age$\rangle$ (Gyr)} & 
\colhead{$\Delta t$ (Gyr)} & \colhead{log Z/Z$_\odot$} & \colhead{E(B--V)} & \colhead{log\,M$_s$/M$_\odot$} & 
\colhead{$\chi^2_{\rm r,min}$}}
\startdata
\multicolumn{9}{c}{Calzetti Reddening \citep{calz01}}\\
SSP & BC07 & 0.38\asec & 0.77$_{-0.06}^{+0.06}$ & {\hfill ---\hfill} & $-$0.58$_{-0.08}^{+0.04}$ & $<0.02$ & 11.33$_{-0.04}^{+0.04}$ & 1.36\\
SSP & BC07 & 0.64\asec & 0.88$_{-0.14}^{+0.07}$ & {\hfill ---\hfill} & $-$0.60$_{-0.07}^{+0.06}$ & $<0.02$ & 11.36$_{-0.06}^{+0.07}$ & 1.45\\
SSP & CM05 & 0.38\asec & 0.77$_{-0.05}^{+0.06}$ & {\hfill ---\hfill} & $-$1.33$_{-0.03}^{+0.07}$ & $<0.02$ & 10.69$_{-0.04}^{+0.04}$ & 3.31\\
SSP & CM05 & 0.64\asec & 0.80$_{-0.06}^{+0.06}$ & {\hfill ---\hfill} & $-$1.21$_{-0.04}^{+0.08}$ & $<0.02$ & 10.69$_{-0.04}^{+0.04}$ & 3.58\\
EXP & BC07 & 0.38\asec & 0.63$_{-0.11}^{+0.10}$ &  0.17$_{-0.03}^{+0.02}$ & $+$0.21$_{-0.08}^{+0.03}$ &  0.06$_{-0.05}^{+0.04}$ & 11.45$_{-0.03}^{+0.03}$ & 1.02\\
EXP & BC07 & 0.64\asec & 0.67$_{-0.07}^{+0.11}$ &  0.17$_{-0.01}^{+0.03}$ & $+$0.20$_{-0.10}^{+0.03}$ &  0.05$_{-0.04}^{+0.03}$ & 11.45$_{-0.04}^{+0.07}$ & 1.11\\
EXP & CM05 & 0.38\asec & 0.45$_{-0.10}^{+0.05}$ &  0.12$_{-0.02}^{+0.03}$ & $+$0.21$_{-0.08}^{+0.03}$ &  $<$0.04              & 10.74$_{-0.03}^{+0.03}$ & 1.78\\
EXP & CM05 & 0.64\asec & 0.47$_{-0.05}^{+0.19}$ &  0.15$_{-0.02}^{+0.04}$ & $+$0.20$_{-0.12}^{+0.04}$ &  0.02$_{-0.01}^{+0.02}$ & 10.75$_{-0.04}^{+0.08}$ & 2.10\\
CST & BC07 & 0.38\asec & 1.44$_{-0.24}^{+0.10}$ &  2.71$_{-0.41}^{+0.12}$ & $+$0.21$_{-0.03}^{+0.03}$ &  0.05$_{-0.01}^{+0.02}$ & 11.59$_{-0.07}^{+0.04}$ & 0.88\\
CST & BC07 & 0.64\asec & 1.46$_{-0.17}^{+0.08}$ &  2.76$_{-0.37}^{+0.07}$ & $+$0.21$_{-0.05}^{+0.03}$ &  0.06$_{-0.01}^{+0.02}$ & 11.60$_{-0.03}^{+0.03}$ & 1.03\\
CST & CM05 & 0.38\asec & 0.73$_{-0.15}^{+0.12}$ &  1.43$_{-0.29}^{+0.08}$ & $+$0.21$_{-0.03}^{+0.03}$ &  0.03$_{-0.02}^{+0.02}$ & 10.82$_{-0.03}^{+0.03}$ & 1.57\\
CST & CM05 & 0.64\asec & 0.78$_{-0.15}^{+0.11}$ &  1.47$_{-0.11}^{+0.26}$ & $+$0.21$_{-0.06}^{+0.03}$ &  0.05$_{-0.01}^{+0.02}$ & 10.82$_{-0.04}^{+0.06}$ & 1.86\\
\hline
\\
\multicolumn{9}{c}{Milky Way Reddening \citep{fitz99}}\\
SSP & BC07 & 0.38\asec & 0.77$_{-0.06}^{+0.06}$ & {\hfill ---\hfill} & $-$0.58$_{-0.08}^{+0.04}$ & $<0.02$ & 11.33$_{-0.04}^{+0.04}$ & 1.36\\
SSP & BC07 & 0.64\asec & 0.89$_{-0.09}^{+0.06}$ & {\hfill ---\hfill} & $-$0.60$_{-0.07}^{+0.06}$ & $<0.02$ & 11.36$_{-0.06}^{+0.07}$ & 1.45\\
SSP & CM05 & 0.38\asec & 0.77$_{-0.06}^{+0.06}$ & {\hfill ---\hfill} & $-$1.32$_{-0.03}^{+0.08}$ & $<0.02$ & 10.68$_{-0.04}^{+0.04}$ & 3.31\\
SSP & CM05 & 0.64\asec & 0.81$_{-0.06}^{+0.06}$ & {\hfill ---\hfill} & $-$1.21$_{-0.04}^{+0.08}$ & $<0.02$ & 10.68$_{-0.04}^{+0.04}$ & 3.58\\
EXP & BC07 & 0.38\asec & 0.63$_{-0.11}^{+0.11}$ &  0.17$_{-0.03}^{+0.01}$ & $+$0.21$_{-0.07}^{+0.03}$ &  0.02$_{-0.02}^{+0.03}$ & 11.44$_{-0.03}^{+0.03}$ & 1.02\\
EXP & BC07 & 0.64\asec & 0.72$_{-0.10}^{+0.09}$ &  0.17$_{-0.02}^{+0.03}$ & $+$0.20$_{-0.13}^{+0.03}$ &  0.02$_{-0.02}^{+0.03}$ & 11.44$_{-0.03}^{+0.03}$ & 1.14\\
EXP & CM05 & 0.38\asec & 0.46$_{-0.04}^{+0.04}$ &  0.13$_{-0.02}^{+0.03}$ & $+$0.21$_{-0.08}^{+0.03}$ &  $<$0.02              & 10.74$_{-0.03}^{+0.03}$ & 1.78\\
EXP & CM05 & 0.64\asec & 0.50$_{-0.08}^{+0.26}$ &  0.16$_{-0.02}^{+0.05}$ & $+$0.12$_{-0.08}^{+0.11}$ &  0.02$_{-0.02}^{+0.01}$ & 10.77$_{-0.06}^{+0.07}$ & 2.13\\
CST & BC07 & 0.38\asec & 1.39$_{-0.23}^{+0.14}$ &  2.46$_{-0.33}^{+0.36}$ & $+$0.21$_{-0.03}^{+0.03}$ &  0.03$_{-0.02}^{+0.01}$ & 11.55$_{-0.06}^{+0.06}$ & 0.92\\
CST & BC07 & 0.64\asec & 1.44$_{-0.27}^{+0.10}$ &  2.68$_{-0.51}^{+0.15}$ & $+$0.15$_{-0.64}^{+0.05}$ &  0.06$_{-0.06}^{+0.01}$ & 11.58$_{-0.08}^{+0.04}$ & 1.09\\
CST & CM05 & 0.38\asec & 0.78$_{-0.15}^{+0.07}$ &  1.46$_{-0.12}^{+0.05}$ & $+$0.21$_{-0.06}^{+0.03}$ &  0.02$_{-0.01}^{+0.02}$ & 10.81$_{-0.03}^{+0.03}$ & 1.60\\
CST & CM05 & 0.64\asec & 1.25$_{-0.50}^{+0.26}$ &  2.56$_{-1.13}^{+0.26}$ & $+$0.21$_{-0.18}^{+0.03}$ &  $<$0.04              & 10.89$_{-0.08}^{+0.09}$ & 1.97\\
\enddata
\label{tab:ModelPreds}
\end{deluxetable*}

\begin{figure*}
\begin{center}
  \includegraphics[width=8cm]{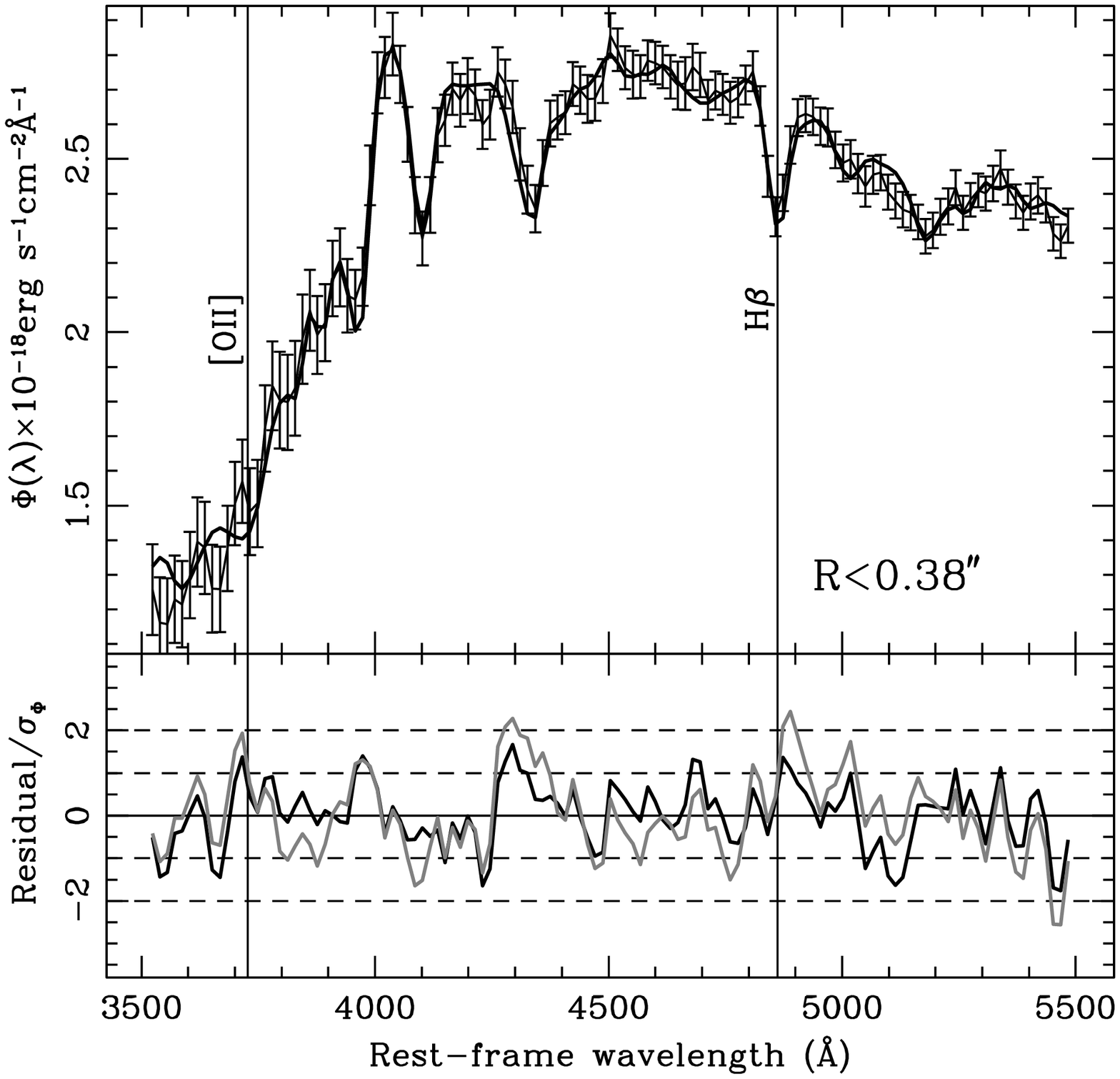}
  \includegraphics[width=8cm]{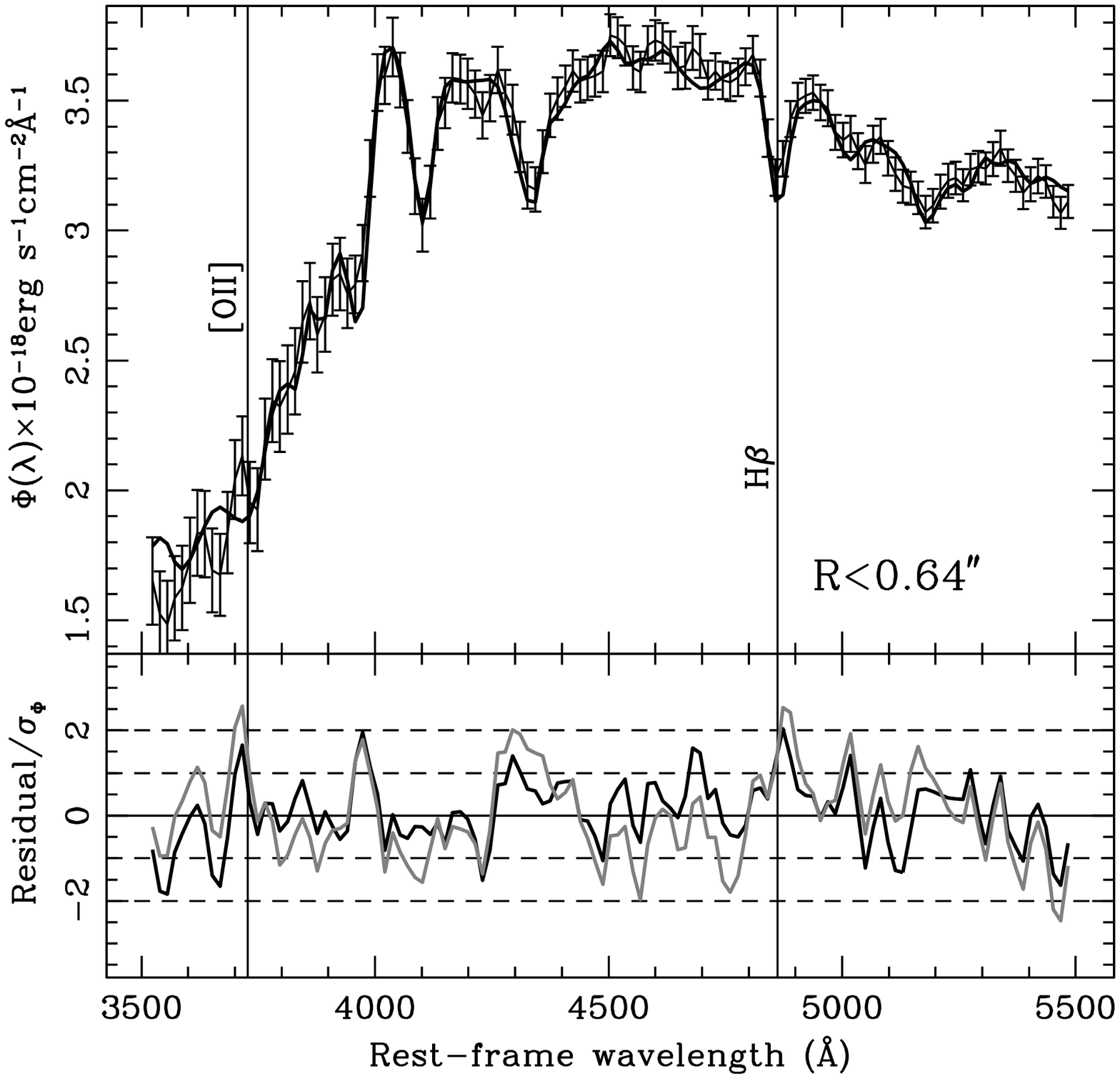}
  \caption{Best fit of the WFC3 slitless grism data for the CST2 models
    shown in the rest-frame, with each panel showing a different
    extraction, as labelled. The residuals (as a fraction of the RMS
    per wavelength bin) are shown in the bottom panel. The black
    (grey) lines correspond to the fits using BC07 (CM05) models.}
  \label{fig:cstsed}
\end{center}
\end{figure*}

\subsection{Exponential Models}

The lack of signatures of ongoing star formation in the SED of FW4871
argues against an exponentially decaying star formation rate. However,
these models have been commonly applied to evolved stellar
populations, and so, we present as an exercise the results of fitting
such models (hereafter EXP models), whereby four parameters describe a
star formation history, namely the formation epoch (t$_{\rm FOR}$, the
cosmic epoch when star formation starts), the timescale of the
exponential ($\tau$), the metallicity (Z, kept fixed for all times
within a model), and the dust content, parameterized by the colour
excess E(B--V). These so-called $\tau$-models improve over the SSPs as
they consist of a mixture of stellar ages, giving a more realistic
representation of the real star formation history, with $\tau$
controlling the width of the age distribution. Tab.~\ref{tab:Params}
shows the range explored for these
parameters. Tab.~\ref{tab:ModelPreds} lists the parameter fits,
including the average age and the RMS of the age distribution (using
the likelihood as the probability distribution function of the
parameter considered). The average mass-weighted stellar age according
to EXP models lies between 0.5 and 0.7~Gyr, depending on the
population synthesis library used. We note the age of the Universe at
the redshift of FW4871 is 3.4~Gyr. The main difference between BC07
and CM05 models is the lower value of the age for the
latter. Nevertheless, both population synthesis models agree with a
rather unphysical short-lived burst of formation ($\tau<0.2$~Gyr).  In
constrast to the SSP runs, solar and super-solar abundances are now
allowed, and the best-fit $\chi^2$ values are significantly
reduced. These models would suggest that the bulk of the stellar
populations of FW4871 were formed in an intense and very short-lived
burst started at redshift z$_{\rm F}\sim$2.5--3.

\subsection{Truncated Models}

There is an important limitation with exponentially decaying models:
the tail of the exponential creates a small, but significant amount of
very young stars. The flux from those stars will affect considerably
the spectral fit. Hence, it is necessary -- for spectra without a
significant amount of star formation -- to adopt an additional set of
models, namely those where the star formation is truncated. For
simplicity, we assume that the star formation rate is constant within
an interval between cosmic time t$_{\rm FOR}$ and t$_{\rm FOR}+\Delta$t,
and then switches abruptly to zero (hereafter, we refer to this set as
CST models). This parametrisation allows us to get a more realistic
estimate of the duration of the burst if no young stars are found in
the spectra (as suggested, e.g. by the colour distribution, the SSP
fits of the spectra, or the absence of strong emission lines). Dust is
also included as a free parameter in the same way as for the SSP and
EXP models.  Once again, Tab.~\ref{tab:Params} shows the range
explored for these models. Fig.~\ref{fig:cst} shows the confidence
levels for these models, using the same observational constraints as
for the EXP and the SSP models. The figure gives the constraints on
the parameters controlling the age (top) and metallicity (bottom). The
slanted lines in the top panels represent an average, mass-weighted
age between 0.5 and 2.5 Gyr (thick line), in steps of 0.5~Gyr. Notice
the degeneracy between age and burst duration is reduced for the
analysis with BC07 models (left panels). The CM05 models allow for the
presence of younger stars, but the $\chi^2$ are slightly worse than
for BC07-based models.

The figure suggests an average age about 1.5~Gyr, with a wide age
spread, between 2-3~Gyr, and metallicities above Z$_\odot$/3. The
results do not vary significantly with respect to the choice of the 0.38\asec
($\sim$R$_{\rm e}$/2) or 0.64\asec ($\sim$R$_{\rm e}$) diameter apertures
considered, suggesting rather homogeneous stellar populations.

\begin{deluxetable*}{lcrrrrrrrr}
\tabletypesize{\scriptsize}
\tablewidth{0pt}
\tablecaption{Parameter fits for the CST2 models, combining the
  WFC3/IR SED, the MgUV index (measured from the ACS SED) and F098M,
  F125W and F160W photometry. For ease of comparison with CST models,
  $\langle$Age$\rangle$ is defined here as the average age of the
  older component. Error bars are quoted at the 5--95 percentiles.}
\tablehead{\colhead{PopSyn} & \colhead{Aperture} & \colhead{$\langle$Age$\rangle$ (Gyr)} & 
\colhead{$\Delta t$ (Gyr)} & \colhead{log (t$_Y$/Gyr)} & \colhead{f$_Y$} & \colhead{E$_Y$(B--V)} &
\colhead{log Z/Z$_\odot$} & \colhead{log\,M$_s$/M$_\odot$} & \colhead{$\chi^2_{\rm r,min}$}}
\startdata
BC07 & 0.38\asec & 1.55$_{-0.19}^{+0.25}$ & 2.37$_{-0.75}^{+0.27}$ & $-$0.85$_{-0.30}^{+0.62}$ & 0.06$_{-0.06}^{+0.26}$ & $<$0.21 & $+$0.22$_{-0.08}^{+0.08}$ & 11.47$_{-0.13}^{+0.13}$ & 0.58\\
BC07 & 0.64\asec & 1.76$_{-0.40}^{+0.20}$ & 1.99$_{-0.34}^{+0.56}$ & $-$0.83$_{-0.45}^{+0.09}$ & 0.06$_{-0.06}^{+0.26}$ & $<$0.28 & $+$0.21$_{-0.08}^{+0.08}$ & 11.47$_{-0.13}^{+0.13}$ & 0.65\\
CM05 & 0.38\asec & 1.86$_{-0.50}^{+0.85}$ & 1.15$_{-1.09}^{+1.18}$ & $-$0.82$_{-0.08}^{+0.08}$ & 0.13$_{-0.10}^{+0.10}$ & $<$0.10 & $+$0.15$_{-0.16}^{+0.13}$ & 10.89$_{-0.26}^{+0.14}$ & 0.97 \\
CM05 & 0.64\asec & 1.94$_{-0.21}^{+0.23}$ & 1.72$_{-0.46}^{+0.41}$ & $-$0.82$_{-0.08}^{+0.08}$ & 0.14$_{-0.05}^{+0.09}$ & $<$0.10 & $+$0.18$_{-0.18}^{+0.11}$ & 10.90$_{-0.13}^{+0.13}$ & 1.12\\
\enddata
\label{tab:ModelPreds2}
\end{deluxetable*}

\subsection{Beyond Truncated Models}

One could argue that the single dust screen, adopted for our models so
far, may not take into account a more complex star formation
history. Complex models come at the price of huge volumes of parameter
space.  Furthermore, the inherent degeneracies present in any analysis
of stellar populations from unresolved spectroscopic data lead us to
follow the approach of mapping as much as possible the full volume of
parameter space. Hence, in order to constrain the presence of a
younger population, {\sl with different dust reddening} to the rest
of galaxy, we define a new set of models (CST2) consisting of the
previous one, i.e. constant star formation truncated after some time
lapse $\Delta t$, along with an extra component made up of a young,
simple stellar population. In order to make the parameter search
manageable, we assume a formation epoch for the old component of z$_{\rm
  FOR}$=10 and assume the same metallicity for both components
(although allowed to be a free parameter). Hence, this model is
described by six free parameters (see Tab.~\ref{tab:Params}). The new
parameters describe the age of the young component (t$_Y$), its
contribution in mass (f$_Y$) and the amount of dust in the old and
young components (E$_O$ and E$_Y$). Given the small differences found with
respect to attenuation laws, we will only consider a standard 
\citet{fitz99} law for CST2 models.

Fig.~\ref{fig:cst2} shows the constraints on the parameters mainly
related to the young component (we do not find significant differences
for the parameters controlling the old component, i.e. the equivalent
of Fig.~\ref{fig:cst} for the CST2 models would be very similar). On the
top panels of Fig.~\ref{fig:cst2}, the amount of dust reddening is
shown both for the old (thin lines) and the young component (thick
lines). Tab.~\ref{tab:ModelPreds2} shows the parameter fits over
the 5--95 percentiles. Note the minimum value of the $\chi^2$
decreases -- although we must emphasize that all models explored here
give acceptable fits, so that a Bayesian evidence criterion would not
be capable of discriminating among these models. Nevertheless, the
CST2 models show that a young component could be present, although with
an age of $145_{-70}^{+450}$\,Myr, and a maximum amount of dust
reddening of E(B--V)$<$0.4\,mag (95\% confidence levels). Its
contribution to the total stellar mass budget changes slightly with
respect to the aperture or the population synthesis models used, but
stays roughly below 20\% for all models, with a best-fit value in the
range 5-15\%.

\begin{deluxetable*}{llrlr}
\tabletypesize{\scriptsize}
\tablecaption{List of parameters used in the models explored in this
  paper. A formation time or epoch can be given either by a
  cosmological time (t$_{\rm FOR}$), or by the corresponding redshift
  (z$_{\rm FOR}$). t$_U$ and t$_{10}$ represent the age of the
  Universe at the redshift of FW4871 (z$_{\rm obs}$=1.893), and the
  cosmological time at redshift z=10, respectively.}
\tablewidth{0pt}
\tablehead{\colhead{Model} & \colhead{Parameter} & \colhead{Range} & 
\colhead{Description} & \colhead{No. Models}}
\startdata
SSP & z$_{\rm FOR}$,t$_{\rm FOR}$     & [10,z$_{\rm obs}$] & Formation Epoch (Single Age) & \\
    & log(Z/Z$_\odot$) & [$-$1.5,$+$0.3] & Metallicity & \\
    & E(B--V)  & [0,0.5] & Dust reddening & \\
    &   &   &  & 262,144\\   
\hline
\\
EXP & z$_{\rm FOR}$,t$_{\rm FOR}$ & [10,z$_{\rm obs}$] & Formation Epoch & \\
    & $\log(\tau/{\rm Gyr})$ & [$-$1,$+$0.6] Gyr & Formation Timescale & \\
    & log(Z/Z$_\odot$) & [$-$1.5,$+$0.3] & Metallicity & \\
    & E(B--V)  & [0,0.5] & Dust reddening & \\
    &   &   &  & 1,048,576\\   
\hline
\\
CST & z$_{\rm FOR}$,t$_{\rm FOR}$ & [10,z$_{\rm obs}$] & Formation Epoch & \\
    & $\Delta$t    & [0,$(t_U-t_{10})\sim 3$] Gyr & Burst Duration & \\
    & log(Z/Z$_\odot$) & [$-$1.5,$+$0.3] & Metallicity & \\
    & E(B--V)  & [0,0.5] & Dust reddening & \\
    &   &   &  & 1,048,576\\   
\hline
\\
CST2 & z$_{\rm FOR}$,z$_{\rm FOR}$ & 10 & Formation Epoch (Fixed) & \\
    & $\Delta$t    & [0,$(t_U-t_{10})\sim 3$] Gyr & Burst Duration & \\
    & E$_O$(B--V)  & [0,0.5] & Dust reddening (Old) & \\
    & t$_Y$ & [0.1,2] Gyr & Young Component (SSP) & \\
    & E$_Y$(B--V)  & [0,0.5] & Dust reddening (Young) & \\
    & f$_Y$ & [0,1] & Young Stellar Mass Fraction & \\
    & log(Z/Z$_\odot$) & [$-$1.5,$+$0.3] & Metallicity (Both Components) & \\
    &   &   &  & 2,985,984\\   
\enddata
\label{tab:Params}
\end{deluxetable*}

Fig.~\ref{fig:cstsed} shows the comparison between the observed and
the best fit WFC3 spectra for the BC07 and CM05 models, as black and
grey lines, respectively. As regards emission lines, [\ion{O}{2}] is
only a 1$\sigma$ residual in the 0.38~arcsec extraction, increasing to
2$\sigma$ in the larger aperture. H$\beta$ is more prominent, at the
2-2.5$\sigma$ level in the CM05 case. However, we note that FW4871 has
an X-ray source counterpart \citep{xray} and the implied X-ray
luminosities could not be expected from a starbursting system unless
rates as high as 1,000M$_\odot$/yr are sustained, leaving us with the
option of an active galactic nucleus \citep{vdk} in an otherwise
gas-poor merger. Alternatively, one could consider the X-ray emission
originating from a hot, diffuse halo.  This X-ray halo can be produced
by the conversion of gravitational potential energy into heat. As this
galaxy went through a major growth phase in its past, it is likely
that this process was accompanied by a major growth phase via infall
of material, providing gravitational potential energy. In this way,
star formation could have continued until the halo was heated up to
sufficiently high temperatures. \citet{ko08} find that the fastest
growing halos would also heat the most. In this case, the residuals
found could indeed be related to some (low-level) ongoing star
formation (although note the lack of blue colours in
Fig.~\ref{fig:cgrad}). During the runs we also compute the absolute
luminosity and the U--V colour for each star formation history. It is
worth noting that the best fit models consistently give a luminosity
of M$_V=-24.89\pm 0.07$ and a colour U--V$=1.3\pm 0.06$ (AB, 90\%
confidence levels) regardless of the model or base population
synthesis.

\begin{figure}
  \includegraphics[width=9cm]{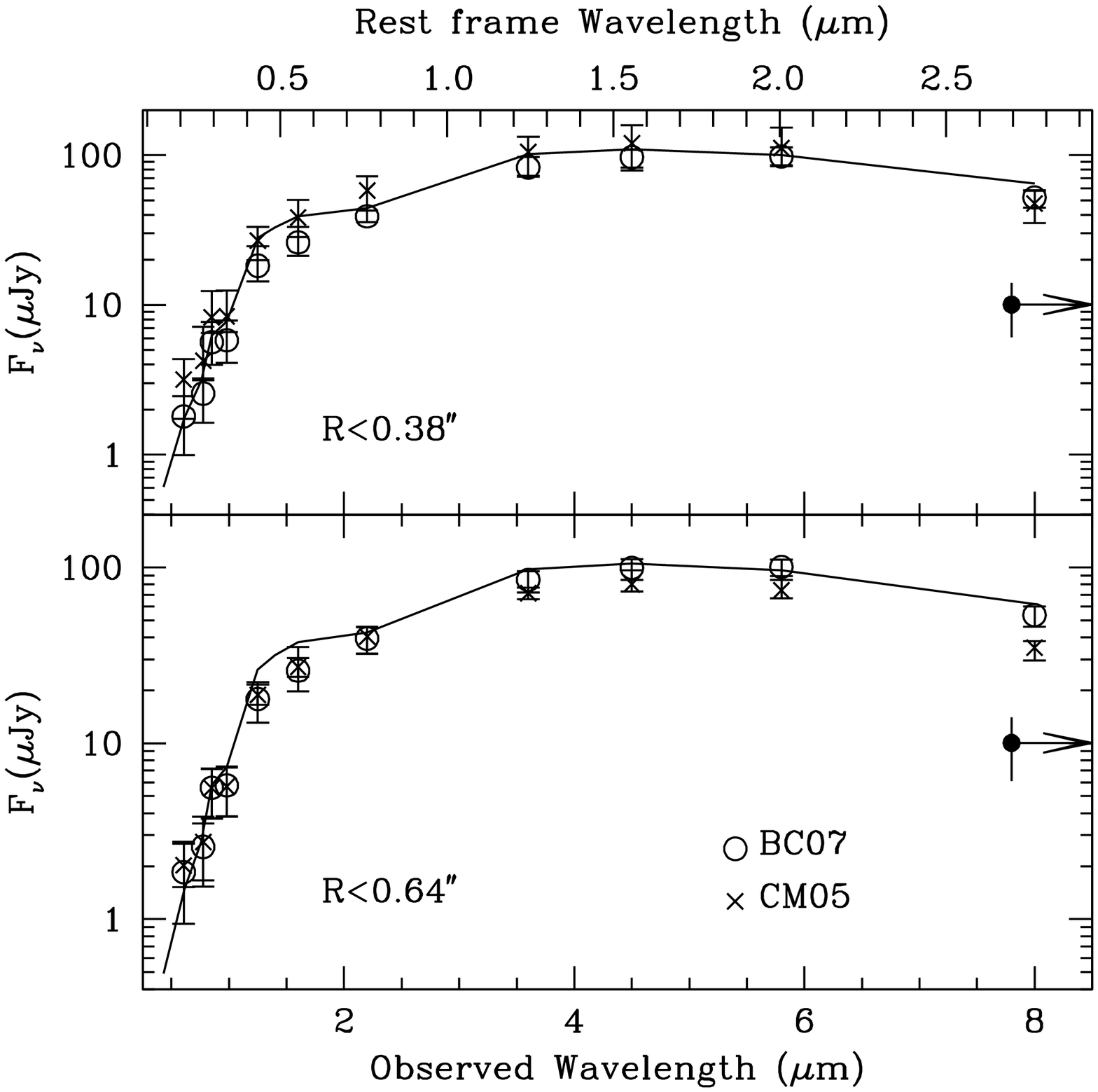}
  \caption{Photometric SED corresponding to the results from the
    spectroscopic fits. The points represent the probability weighted
    fluxes according to the CST2 models, with error bars extending
    over the 90\% confidence interval. The line shows the photometric
    data from HST/ACS, VLT/ISAAC, HST/WFC3-IR and Spitzer/IRAC. The
    solid dot with an arrow marks the observed frame 24$\mu$m flux
    from Spitzer/MIPS \citep[taken from][]{fw}.}
  \label{fig:photsed}
\end{figure}

As a final check of our (mainly spectroscopic) fits with respect to
broadband photometry, covering a wider range of wavelengths, we show
in Fig.~\ref{fig:photsed} a comparison between the observed
photometric data and the best fit CST2 models obtained from spectral
fitting. The observations are given in the figure as a solid line,
whereas the model fits are shown as open circles (BC07) or crosses
(CM05), with error bars representing the 90\% confidence levels. The
observations cover a wide range, from the optical (HST/ACS) to the IR
(Spitzer/IRAC), including NIR (VLT/ISAAC and HST/WFC3-IR). Notice the
constraint from spectral fitting only use data out to $\lambda_{\rm
  obs}\sim 1.6\mu$m, whereas the photometric SED shown here as a solid
line, comes from the total fluxes \citep[taken from the FIREWORKS
  survey,][]{fw}. In the case of the ACS and WFC3 data, we performed
aperture photometry on the images, so that those fluxes correspond to
the chosen 0.38 and 0.64\asec apertures. The IRAC fluxes cannot be
treated this way, given the poorer spatial resolution. The fit is good
for most of the photometric data points with BC07 models. In contrast,
the CM05-based models give slightly higher NUV fluxes and lower
8$\mu$m fluxes, an aspect that cannot be blamed on dust, since the
models give a strong constraint in E(B--V) (see
Tabs.~\ref{tab:ModelPreds} and \ref{tab:ModelPreds2}). An increase in
the amount of dust would drastically worsen the SED fitting of the
WFC3 data, whose superb flux calibration cannot accommodate larger
amounts of reddening. The relatively weak flux observed in the
24$\mu$m Spitzer/MIPS passband \citep[$\Phi_{24}=10.05\,\mu$Jy, compared
  with an 8$\mu$m flux of $\Phi_8=62.0\,\mu$Jy,][]{fw} confirms that
emission from dust -- and thus significant star formation -- is rather
low in this galaxy.

\section{Discussion and Conclusions}

FW4871 constitutes a typical case of a massive galaxy already in place
by redshift z$\sim$2. This type of galaxy is expected to evolve into
the massive early-type galaxies we see at low redshift.  Taking the
redshift catalogue from the ESO database \citep{eso1,eso2} we find 13
(54) galaxies within a redshift range 1.8$<$z$<$2.0 at a projected
separation below $\pm$1 (2) Mpc.  Hence, regarding environment -- over
the large scales mapped by this redshift range -- this galaxy clearly
sits in an overdense region.

We find that a realistic model consisting of a truncated SFH gives no
significant star formation in the recent past history of FW4871,
whereas the duration of the bursting episode extends for around
2.5~Gyr. Unfortunately the best fit values of the likelihood (i.e. the
``minimum $\chi^2$'') is similar among sets of models, preventing us
to discriminate among SSP, EXP, CST, CST2, or other sets of models
using, for instance, a Bayesian evidence estimator, even with better
spectral resolution \citep[see][for a detailed analysis in
  low-redshift galaxies]{bmc}.  One can only resort to prior
information in the sense of rejecting models such as an SSP, where the
formation takes place over a timescale much shorter than the dynamical
timescales of the systems concerned, or EXP models, where a protracted
tail of star formation is at odds with the lack of line emission
related to ongoing star formation. Nevertheless, all models point to
the fact that no significant star formation took place during the
latest stages of FW4871, at least within one effective
radius. Furthermore, the colour profile of this galaxy
(Fig.~\ref{fig:cgrad}) suggests a similar outcome in the external
regions, out to 3R$_{\rm e}$.

We emphasize that within the 4.65~arcmin$^2$ field of view of the
ERS/GOODS data, FW4871 is the only galaxy with a strong Balmer
decrement observed in the redshift range sensitive to the WFC3 grism
data (i.e. 1$<$z$<$3). Given the limit in apparent magnitude of the
observations, this constraint is compatible with the comoving number
density of massive galaxies at z$\sim$2
\citep[$n\sim$10$^{-4}$Mpc$^{-3}$,][]{con07}. Is FW4871 an
outlier or a typical example of a forming massive galaxy?
Larger surveys to a comparable depth are needed to answer this
question. 

The analysis of FW4871 by \citet{vdk} compared the WFC3-IR spectrum
with a small set of synthetic spectra, proposing a stellar age of
0.5~Gyr and A$_V\sim $1~mag, with a tentative, although very weak,
detection of H$\beta$ in emission in the outer regions of the
galaxy. Their spectral fit only presented three possible star
formation histories, without any reference to the range of models
explored. Hence, their suggested scenario cannot include an estimate
of uncertainty in the properties of the underlying stellar
populations. In addition, we note that \citet{vdk} extracted the data
from the publicly available server and performed the reduction of the
data while the latest calibration files were not available. Our team
used the best possible calibration files, obtaining a slightly
different SED compared to the one presented in \citet{vdk}, hence the
difference with respect to the dust content. Even though the simple
scenarios presented in their paper are compatible with our analysis
regarding age, their estimate of dust cannot be accommodated with the
spectrum reduced with the newest calibration data.

We extend the analysis of \citet{vdk} by exploring a wide range of
star formation histories, and include the colour gradients to
strengthen the hypothesis that the star formation in FW4871 has been
truncated in the recent past of its history. The main conclusion from
our analysis of the stellar populations within the effective radius is
that the process of formation started at redshift z$_{\rm F}\sim$10
and lasted for about $\sim$2.5~Gyr.  The addition of an extra,
younger, component to the analysis allows for up to 20\% in mass of a
population with an age of $\sim$150~Myr, with a modest amount of dust
reddening (E(B--V)$<$0.4~mag), therefore with the bulk of stellar
populations having formed at an earlier phase.  The long $\Delta t$ of
the models along with the lack of star forming features in the
spectrum suggests a truncation of the star formation over a few
hundreds of Myr. Taking into account the stellar mass of the galaxy,
we can infer a sustained star formation rate during its bursting phase
in the range 30--110\,M$_\odot$/yr (with the quoted range including
the difference between the stellar masses according to BC07 or CM05
models).  One expects the star-forming progenitors to be similar to
SCUBA galaxies \citep[see e.g.][]{chap05}, at z$\gtrsim$2.5. FW4871
represents the endpoint of the strong star formation activity of a
massive galaxy, perhaps followed by a number of minor mergers -- such
as the one expected with its neighbour FW4887 -- that will increase
its compact size towards those of local massive galaxies
\citep{ks06,naab09}.  Regarding the discrepancy in the stellar mass
given by either BC07 or CM05 models, we note that this disagreement is
caused by the difference in the treatment of evolved phases of stellar
evolution, with TP-AGB stars contributing significantly more to the
luminosity in the CM05 models, thereby reducing the M/L especially for
the ages considered. However, we note that this predicted contribution
still awaits confirmation by observations \citep[e.g.][]{zibetti:12}.

Given the mass of this galaxy, one should expect a major contribution
to its formation via cold accretion at high redshift \citep[see
  e.g.][]{dekbir06}, a process that would boost the star formation
rate both from the enhanced gas accretion rate and a possible increase
of the star formation efficiency \citep[see e.g.][]{ks09}. The derived
star formation history of FW4871 provides evidence -- via the superb
spectroscopic data from HST -- towards a highly efficient channel of
galaxy formation that started around z$_{\rm F}\sim$10, and was
subsequently truncated around z$\gtrsim$2, followed by a brief
quiescent phase. What caused this truncation? Two main scenarios can
be brought forward: i) The onset of a galactic superwind from stellar
feedback: a sustained star formation rate of about
$\sim$100M$_\odot$/yr for a couple of Gyr will provide enough energy;
ii) The previous merging process that resulted in the formation of
FW4871 could be the ultimate cause of the quenching, possibly by
switching of the AGN \citep{ham11}.  The morphological appearance of
FW4871, and the X-ray detection of this source \citep{xray} points to the
latter. In this case, the recent merging process to form FW4871 could
be considered as the cause of truncation. One could envision FW4871 as
a high redshift equivalent of a post-starburst, or k+a galaxy. The
analysis of the environment in a sample of k+a galaxies from the Sloan
Digital Sky Survey \citep{goto05} reveals that the truncation of the
star-bursting phase is caused by the interaction with a nearby
companion, a similar mechanism to the one contemplated here.

This process would make major mergers an important factor to explain
the ``road to the red sequence'', namely the process of quenching of
star formation in massive galaxies, an issue that may require proper
calibration of semi-analytic models of galaxy formation.

\acknowledgments 
We gratefully acknowledge enlightening discussions with Daisuke Kawata
and Myrto Symeonidis. We made extensive use of the delos computer
cluster in the physics department at King's College London, with
thanks to the sysadmin Dr. Nigel Arnot.

Facilities: \facility{\hst\ (ACS,WFC3)}



\begin{thebibliography}{}

\bibitem[\protect\citeauthoryear{Baldry et al.}{2004}]{bal04}
Baldry, I.~K., Glazebrook, K., Brinkmann, J., \u{Z}eljko, I., Lupton,
R.~H., Nichol, R.~C., Szalay, A.~S., 2004, ApJ, 600, 681

\bibitem[\protect\citeauthoryear{Balogh et al.}{1999}]{bal99}
Balogh, M.~L., Morris, S.~L., Yee, H.~K.~C., Carlberg, R.~G., 
Ellingson, E.,1999, ApJ, 527, 54

\bibitem[\protect\citeauthoryear{Banerji et al.}{2010}]{ban10}
Banerji, M., Ferreras, I., Abdalla, F., Hewet, P., Lahav, O., 2010, MNRAS, 402, 2264

\bibitem[\protect\citeauthoryear{Balestra et al.}{2010}]{eso2}
Balestra, I., et al. 2010, A\& A, 512, 12

\bibitem[\protect\citeauthoryear{Bournaud et al.}{2007}]{bour07}
Bournaud, F., Jog, C.~J., Combes, F., 2007, A\& A, 476, 1179

\bibitem[\protect\citeauthoryear{Bruzual \& Charlot}{2003}]{bc03}
Bruzual, G., \& Charlot, S., 2003, MNRAS, 344, 1000

\bibitem[\protect\citeauthoryear{Bruzual}{2007}]{bruz07}
  Bruzual, G. 2007, IAU No. 241 Symp. Procs.
  "Stellar populations as building blocks of galaxies", 
  eds. A. Vazdekis and R.~F. Peletier, Cambridge,
  arXiv:astro-ph/0703052

\bibitem[\protect\citeauthoryear{Calzetti}{2001}]{calz01}
Calzetti, D., 2001, PASP, 113, 1449

\bibitem[\protect\citeauthoryear{Chapman et al.}{2005}]{chap05}
Chapman, S.~C., Blain, A.~W., Smail, I., Ivison, R.~J., 2005, ApJ,
622, 772

\bibitem[\protect\citeauthoryear{Cimatti et al.}{2008}]{cim08}
Cimatti, A., et al., 2008, A\& A, 482, 21

\bibitem[\protect\citeauthoryear{Conroy et al.}{2010}]{bump}
Conroy, C., Schiminovich, D., Blanton, M.~R., 2010, ApJ, 718, 184

\bibitem[\protect\citeauthoryear{Conselice et al.}{2007}]{con07}
Conselice, C.~J., et al. 2007, MNRAS, 381, 962

\bibitem[\protect\citeauthoryear{Daddi et al.}{2005}]{dad05}
Daddi, E., et al., 2005, ApJ, 626, 680

\bibitem[\protect\citeauthoryear{Damjanov et al.}{2009}]{dam09} 
Damjanov, I., et al. 2009, ApJ, 695, 101

\bibitem[\protect\citeauthoryear{Dekel \& Birnboim}{2006}]{dekbir06}
Dekel, A., Birnboim, Y., 2006, MNRAS, 368, 2

\bibitem[\protect\citeauthoryear{Dekel et al.}{2009}]{dek09}
Dekel, A., et al. 2009, Nature, 457, 451

\bibitem[\protect\citeauthoryear{Dressler \& Gunn}{1992}]{dg92} 
Dressler, A. \& Gunn, J.~E., 1992, ApJS, 78, 1

\bibitem[\protect\citeauthoryear{Dunkley et al.}{2009}]{dunk09}
Dunkley, J., et al. 2009, ApJS, 180, 306

\bibitem[\protect\citeauthoryear{Ferreras \& Yi}{2004}]{ig04}
Ferreras, I., Yi, S.~K., 2004, MNRAS, 350, 1322

\bibitem[\protect\citeauthoryear{Ferreras et al.}{2009}]{pears}
Ferreras, I., Pasquali, A., et al., 2009, ApJ, 706, 158

\bibitem[\protect\citeauthoryear{Ferreras et al.}{2005}]{ig05}
Ferreras, I., Lisker, T., Carollo, C.~M., Lilly, S.~J., Mobasher, B.,
2005, ApJ, 635, 243

\bibitem[\protect\citeauthoryear{Ferreras et al.}{2009}]{ig09}
Ferreras, I., Lisker, T., Pasquali, A., Khochfar, S., Kaviraj, S.,
2009, MNRAS, 396, 1573

\bibitem[\protect\citeauthoryear{Fitzpatrick}{1999}]{fitz99}
Fitzpatrick, E.~L., 1999, PASP, 111, 63

\bibitem[\protect\citeauthoryear{Fan et al.}{2008}]{fan08}
Fan, L., Lapi, A., De Zotti, G., Danese, L. 2008, ApJ, 689, L101

\bibitem[\protect\citeauthoryear{Fontana et al.}{2006}]{fon06}
Fontana, A., et al. 2006, A\& A, 459, 745

\bibitem[\protect\citeauthoryear{Goto}{2005}]{goto05}
Goto, T., 2005, MNRAS, 357, 937

\bibitem[\protect\citeauthoryear{Hambrick et al.}{2011}]{ham11}
Hambrick, D.~C., Ostriker, J.~P., Naab, T., 
Johansson, P.~H., 2011, ApJ, 738, 16

\bibitem[\protect\citeauthoryear{Kauffmann et al.}{2003}]{kauf03}
Kauffmann, G., et al. 2003, MNRAS, 341, 54

\bibitem[\protect\citeauthoryear{Kere\u{s} et al.}{2009}]{keres09}
Kere\u{s}, D., Katz, N., Fardal, M., Dav\'e, R., Weinberg, D.~H., 2009,
MNRAS, 395, 160

\bibitem[\protect\citeauthoryear{Khochfar \& Silk}{2006}]{ks06}
Khochfar, S., Silk, J., 2006, MNRAS, 370, 902

\bibitem[\protect\citeauthoryear{Khochfar \& Silk}{2006b}]{ks06b}
Khochfar, S., Silk, J., 2006b, ApJ, 648, L21

\bibitem[\protect\citeauthoryear{Khochfar \& Ostriker}{2008}]{ko08}
Khochfar, S. \& Ostriker, J.~P., 2008, ApJ, 680, 54

\bibitem[\protect\citeauthoryear{Khochfar \& Silk}{2009}]{ks09}
Khochfar, S., Silk, J., 2009, ApJ, 700, L21

\bibitem[\protect\citeauthoryear{Komatsu et al.}{2011}]{kom11}
Komatsu, E., et al., 2011, ApJS, 192, 18

\bibitem[\protect\citeauthoryear{Kriek et al.}{2009}]{kriek09}
Kriek, M., van Dokkum, P.~G., Labb\'e, I., Franx, M., Illingworth,
G.~D., Marchesini, D., Quadri, R.~F., 2009, ApJ, 700, 221 

\bibitem[\protect\citeauthoryear{K\"ummel et al.}{2009}]{mk09} 
K\"ummel, M., Walsh, J.~R., Pirzkal, N., Kuntschner, H., Pasquali, A., 2009,
PASP, 121, 59 

\bibitem[\protect\citeauthoryear{K\"ummel et al.}{2010}]{mk11} 
K\"ummel, M., Kuntschner, H., Walsh, J.~R., Bushouse, H., 2011, ASPC, 442, 525

\bibitem[\protect\citeauthoryear{Longhetti et al.}{2007}]{lon07}	
Longhetti, M., et al. 2007, MNRAS, 374, 614

\bibitem[\protect\citeauthoryear{Lucy}{1974}]{lucy74}	
Lucy, L.~B., 1974, AJ, 79, 745

\bibitem[\protect\citeauthoryear{Luo et al.}{2008}]{xray}	
Luo, B,  et al. 2008, ApJS, 179, 19

\bibitem[\protect\citeauthoryear{Maraston}{2005}]{cm05}
Maraston, C., 2005, MNRAS, 362, 799

\bibitem[\protect\citeauthoryear{Maraston et al.}{2006}]{cm06}
Maraston, C., Daddi, E., Renzini, A., Cimatti, A., 
Dickinson, M., Papovich, C., Pasquali, A., Pirzkal, N., 2006, ApJ, 652, 85

\bibitem[\protect\citeauthoryear{Naab et al.}{2009}]{naab09}
Naab, T., Johansson, P.~H., Ostriker, J.~P., 2009, ApJ, 699, L178

\bibitem[\protect\citeauthoryear{Nair et al.}{2011}]{na11}
Nair, P., van den Bergh, S., Abraham, R.~G., 2011, ApJ, 734, L31

\bibitem[\protect\citeauthoryear{Pasquali et al.}{2006}]{grapes}
Pasquali, A., et al., 2006, ApJ, 636, 115

\bibitem[\protect\citeauthoryear{Peng et al.}{2010}]{galfit}
Peng, C.~Y., Ho, L.~C., Impey, C.~D., Rix, H.-W., 2010, AJ, 139, 2097

\bibitem[\protect\citeauthoryear{P\'erez-Gonz\'alez et al.}{2008}]{pg08}
P\'erez-Gonz\'alez, P.~G., Trujillo, I., Barrp, G., Gallego, J.,
Zamorano, J., Conselice, C.~J., 2008, ApJ, 687, 50

\bibitem[\protect\citeauthoryear{Popesso et al.}{2009}]{eso1}
Popesso, P., et al., 2009, A\& A, 494, 443

\bibitem[\protect\citeauthoryear{Ragone-Figueroa \& Granato}{2011}]{rag11}
Ragone-Figueroa, C., Granato, G.~L., 2011, MNRAS, 414, 3690

\bibitem[\protect\citeauthoryear{Riess et al.}{2011}]{riess11}
Riess, A., et al. 2011, ApJ, 730, 119 

\bibitem[\protect\citeauthoryear{Rogers et al.}{2010}]{bmc}
Rogers, B., Ferreras, I., Peletier, R., Silk, J., 2010, MNRAS, 402, 447

\bibitem[\protect\citeauthoryear{Saglia et al.}{2000}]{sag00}
Saglia, R.~P., Maraston, C., Greggio, L., Bender, R., Ziegler, B.,
2000, A\& A, 360, 911 

\bibitem[\protect\citeauthoryear{Shankar et al.}{2010}]{shan10}
Shankar, F., Marulli, F., Bernardi, M., Dai, X., Hyde, J.~B., Sheth,
R.~K., 2010, MNRAS, 403, 117

\bibitem[\protect\citeauthoryear{Spinrad et al.}{1997}]{spin97}
Spinrad, H., Dey, A., Stern, D., Dunlop, J., Peacock, J., 
Jimenez, R., Windhorst, R., 1997, ApJ, 484, 581

\bibitem[\protect\citeauthoryear{Straughn et al.}{2011}]{stra11}
Straughn, A.~N., et al., 2011, AJ, 141, 14  

\bibitem[\protect\citeauthoryear{Trujillo et al.}{2006}]{truj06}
Trujillo, I., et al., 2006, ApJ, 650, 18

\bibitem[\protect\citeauthoryear{Trujillo et al.}{2011}]{i3}
Trujillo, I., Ferreras, I., de la Rosa, I.~G., 2011, MNRAS, 415, 3903

\bibitem[\protect\citeauthoryear{van der Wel et al.}{2009}]{vdw09}
van der Wel, A., Bell, E.~F., van den Bosch, F.~C., Gallazzi, A., Rix,
H.-W., 2009, ApJ, 698, 1232

\bibitem[\protect\citeauthoryear{van Dokkum et al.}{2008}]{vdk08}
van Dokkum, P.~G., et al. 2008, ApJ, 677, L5

\bibitem[\protect\citeauthoryear{van Dokkum \& Brammer}{2010}]{vdk}
van Dokkum, P.~G., Brammer, G., 2010, ApJ, 718, L73

\bibitem[\protect\citeauthoryear{White \& Rees}{1978}]{wr78}
White, S.~D.~M., Rees, M.~J., 1978, MNRAS, 183, 341

\bibitem[\protect\citeauthoryear{Windhorst et al.}{2011}]{ers}
Windhorst, R.~A., et al. 2011, ApJS, 193, 27

\bibitem[\protect\citeauthoryear{Wuyts et al.}{2008}]{fw}
Wuyts, S., Labb\'e, I., Schreiber, N.~M,~F., Franx, M., Rudnick, G., 
Brammer, G.~B., van Dokkum, P.~G,. 2008, ApJ, 682. 985 

\bibitem[\protect\citeauthoryear{Zibetti et al.}{2012}]{zibetti:12}
Zibetti, S., Gallazzi, A., Charlot, S,. Pasquali, A., Pierini, D.,
2012, Proc. of IAU Symp. 284, ``The Spectral Energy Distribution of Galaxies'', R.~J. Tuffs, C.~C. Popescu, eds., arXiv:1203.4571

  
\end{thebibliography}
\end{document}